\documentclass[aip,amsmath,amssymb]{revtex4-1}
\usepackage{graphicx}
\usepackage{dcolumn}
\usepackage{bm}
\usepackage{tabularx}
\draft 
\usepackage{amsmath}
\DeclareUnicodeCharacter{2064}{}

\usepackage[version=4]{mhchem} 
\usepackage{booktabs}
\usepackage{siunitx}
\DeclareSIUnit[number-unit-product = {\,}]\kcal{kcal}

\usepackage{color}

\newcommand{\BR}[1]{{\color{red} #1}}

\begin{document}

\date{\today}

\title{Quadrupolar NMR Relaxation as a Local Probe of Collective Dynamics in Aqueous Alcaline and Alcaline-Earth Chlorides Solutions}

\author{Matthieu Wolf}
\affiliation{Sorbonne Universit\'e, CNRS, Physicochimie des \'Electrolytes et Nanosyst\`emes Interfaciaux, F-75005 Paris, France}
\author{Iurii Chubak}
\affiliation{Sorbonne Universit\'e, CNRS, Physicochimie des \'Electrolytes et Nanosyst\`emes Interfaciaux, F-75005 Paris, France}
\author{Benjamin Rotenberg}
\affiliation{Sorbonne Universit\'e, CNRS, Physicochimie des \'Electrolytes et Nanosyst\`emes Interfaciaux, F-75005 Paris, France}
\affiliation{R\'eseau sur le Stockage Electrochimique de l'Energie (RS2E), FR CNRS 3459, 80039 Amiens Cedex, France}
\email{benjamin.rotenberg@sorbonne-universite.fr}



\begin{abstract}
While nuclear magnetic resonance (NMR) provides valuable insights into the local environment of many nuclei, the unambiguous interpretation of the signal in terms of microscopic dynamics is often difficult, particularly when the quadrupolar relaxation mechanism comes into play. Here, we investigate the quadrupolar NMR relaxation of cations and anions in aqueous alcaline and alcaline-earth chlorides solutions, across a broad range of salt concentrations. Using a combination of DFT calculations and classical molecular dynamics simulations, we compute the electric field gradient (EFG) fluctuations over the relevant time scales. Predicted NMR relaxation rates are in good agreement with experiments from the literature. As previously reported for NaCl, we find that the increase in relaxation rate with salt concentration is primarily driven by the slowing of EFG fluctuations, while changes in the static variance of the EFG play a minor role. We highlight some specific features for smaller and divalent cations compared to the other monovalent ones. Additionally, we assess the relevance of the Stokes-Einstein-Debye model, frequently used to analyze NMR relaxation experiments, for these aqueous electrolytes, and highlight the link between the collective dynamics of the liquid underlying the EFG fluctuations at the ion positions and the stress fluctuations. Our results generalize observations for Na$^+$ in aqueous NaCl solutions, showing that models assuming a viscous model of the solvent dynamics are insufficient to describe EFG fluctuations in these systems and illustrate the relevance of molecular simulations to interpret NMR relaxation experiments in terms of microscopic dynamics.
\end{abstract}

\maketitle

\newpage

\section{Introduction}

Detailed information on the solvation shell structure and dynamics of aqueous alkaline and  alkaline-earth cations can shed light on multiple technological, chemical, and biological phenomena~\cite{Ohtaki1993,Madelin2014,madelin2022x,Chandrashekar2012,Pecher2017}. While nuclear magnetic resonance (NMR) can provide valuable insights on chemical arrangements and relaxation in the environment of many nuclei, the unambiguous interpretation of the signal is often hindered when the quadrupolar relaxation mechanism comes into play. The latter applies to nuclei with spin $I \geq 1$ (such as $^7{\rm Li}^+$, $^{23}{\rm Na}^+$, $^{39}{\rm K}^+$, $^{25}{\rm Mg}^{2+}$, $^{43}{\rm Ca}^{2+}$, $^{133}{\rm Cs}^{+}$, $^{85}{\rm Rb}^{+}$, etc), for which the main interaction is that between the nuclear quadrupolar moment $eQ$ and the electric field gradient (EFG) tensor at its position \cite{abragam1961principles}.

The strength of the quadrupolar interaction together with thermal fluctuations of molecular environemnts of the nucleus (having a charactericstic correlation time $\tau_{\rm c}$) govern the relaxation of the distribution of nuclear spins in the external static magnetic field $B_0$ towards equilibrium~\cite{abragam1961principles}. Assuming that molecular processes are sufficiently fast, \textit{i.e} a Larmor frequency $\gamma_n B_0 \ll \tau_{\rm c}^{-1}$ with $\gamma_n$ the nuclear gyromagnetic ratio (so-called extreme narrowing regime), the spin-lattice $1/T_1$ and spin-spin $1/T_2$ rates describing the magnetization relaxation along the longitudinal and transverse directions with respect to $B_0$, respectively, become equal. They further read~\cite{abragam1961principles,RS1993}:
\begin{equation}
\label{eq:nmr_rates}
\frac{1}{T_1} = \frac{1}{T_2} = \frac{1}{20}
\frac{2I+3}{I^2(2I-1)} \left(\frac{eQ}{\hbar}\right)^2 
\int_0^{\infty} {\rm d}t \,
\left\langle {\mathbf{V}}(t)\,{:}{\mathbf{V}}(0) \right\rangle,
\end{equation}
where ${\mathbf{V}}$ is the full EFG tensor at the nucleus that includes the electronic cloud contribution, $C_{\rm EFG}(t) \equiv \left\langle {\mathbf{V}}(t)\,{:}{\mathbf{V}}(0) \right\rangle =
\left\langle \sum_{\alpha,\beta} V_{\alpha\beta}(t) V_{\alpha\beta}(0) \right\rangle$
($\alpha, \beta = x, y, z$) is the EFG autocorrelation function (ACF), and
$\hbar$ stands for the reduced Planck constant. This expression, obtained within linear response theory, allows to predict the relaxation time from equilibrium simulations in the absence of applied magnetic field: $\langle \dots \rangle$ denotes an average in the canonical ensemble, with fixed number of particles $N$, system volume $V$ and temperature $T$.

While quadrupolar NMR relaxation rates \eqref{eq:nmr_rates} can elucidate the structure and dynamics of ionic solvation shells, the interpretation of experimental data is often challenging due to the limited knowledge of quadrupolar coupling constants and of molecular mechanisms behind relaxation. A series of models have thus been proposed to rationalize the observations. Hynes and Wolynes developed an analytical theory based on the quadrupole reorientation dynamics in a continuous dielectric solvent~\cite{Hynes1981}. Perng and Ladanyi dropped the continuous solvent description in their refined dielectric theory coupling the field gradient dynamics to solvent charge density fluctuations~\cite{Perng1998}. While the continuous dielectric theory of Hynes and Wolynes~\cite{Hynes1981} typically underestimated the predicted values of the quadrupolar rates, the theory of Perng and Ladanyi~\cite{Perng1998} was able to achieve quantitative agreement with experiments, yet it relied on a free parameter. Bosse \emph{et al.} formulated a mode-coupling theory for the EFG relaxation in molten salts~\cite{Bosse1983}. A series of models by Hertz \emph{et al.} related the correlation time of EFG fluctuations to water dipole reorientation~\cite{Hertz1973_1, Hertz1973_2, Hertz1974}. However, this molecular mechanism was shown to oversimplify many-body correlations in the context of quadrupolar relaxation dynamics~\cite{Versmold1986, RS1993, Carof2016}. More recently, the EFG correlation time was suggested to be interpreted as the rotational correlation time for an object in a continuous solvent \cite{Price1990,Mitchell2016,DAgostino2021}, likely to be associated with collective stochastic rotations of the ionic solvation shell~\cite{Eisenstadt1966, Eisenstadt1967}. While rotational diffusion is clearly at play in the case of \emph{intra}molecular quadrupolar relaxation~\cite{Abragam1961}, as for the relaxation of \textsuperscript{2}H in heavy water, its applicability to the problem of single quadrupolar ions that is inherently \emph{inter}molecular requires a thorough investigation. 

The predictions of the above-mentioned theories have been tested both with ab initio~\cite{Badu2013, Schmidt2008, Philips2017, Philips2018, Philips2020} and classical~\cite{EngJon1982, EngJon1984, Linse1989, RS1993, Carof2014, Carof2015, Carof2016, Mohammadi2020Nov, Chubak2021Oct, Chubak2023Jan} MD simulations. In particular, the isotropic monoexponential character of the quadrupolar relaxation dynamics that is often assumed in theories with a continuous solvent description was challenged both at the ab initio and classical levels. Using classical MD, Roberts and Schnitker~\cite{RS1993} highlighted a pronounced effect of intermolecular cross correlations on the EFG relaxation that are treated in a rather simplified fashion within molecular dipole reorientation models \cite{Hertz1973_1, Hertz1973_2, Hertz1974}. Carof \emph{et al.}~\cite{Carof2016} underlined a major role of collective fluctuations in ionic solvation shells on the EFG dynamics. 

Compared to classical MD approaches that rely on the Sternheimer approximation for the electron cloud contribution to the EFG at the ion position~\cite{Sternheimer1950, Sternheimer1966, Carof2014, Chubak2021Oct, Chubak2023Jan}, ab initio methods, in particular those relying on density functional theory (DFT) calculations~\cite{Autschbach2010}, provide a very good accuracy of the computed EFGs \cite{Bloechl1994,Autschbach2010,Charpentier2011,Badu2013,Philips2017,Philips2020}. Nevertheless, the high cost of such simulations limits the sampling of the EFG fluctuations~\cite{Badu2013,Schmidt2008,Philips2017} over the time scales necessary to precisely compute the integral in Eq.~\eqref{eq:nmr_rates}. Therefore, on one hand, the use of the fully first-principles approach may not be appropriate in concentrated electrolyte solutions considered here, for which the quadrupolar NMR relaxation slows down with increasing salt concentration and decreasing temperature \cite{Hertz1973_1, Hertz1973_2, Hertz1974, Eisenstadt1966, Eisenstadt1967, Chubak2023Jan}. On the other hand, the Sternheimer factors used in classical MD approaches were shown to be sensitive to the local charge distribution around the ion generated with a specific force field at hand~\cite{Carof2014, Chubak2021Oct}, and their concentration dependence might be necessary to be taken into account~\cite{Chubak2023Jan}. Thus, while classical MD offers a better precision in determining the statistical relaxation of the field gradient fluctuations, it must be coupled with pertinent estimates of the Sternheimer factors to describe local chemical environments.

Here, building on our previously developed strategy, we combine DFT calculations with classical molecular dynamics simulations to investigate the concentration dependence of the quadrupolar NMR relaxation rates of ions in aqueous alcaline and alcaline-earth chloride solutions and their relation to solution viscosity. The systems and methods are presented in Section~\ref{sec:DATA-AND-METHODS}, while results are shown and discussed in Section~\ref{sec:RESULTS}.


\section{System and methods}
\label{sec:DATA-AND-METHODS}

\subsection{Molecular dynamics}
\label{sec:MD}

We performed classical molecular dynamics simulations of a set of aqueous electrolyte solutions (LiCl, NaCl, KCl, RbCl, CsCl, MgCl$_{2}$ and CaCl$_{2}$) at salt molalities ranging from $c = 0.06$ to 4 m, with LiCl extended up to 10 m, using the Madrid-2019 force field~\cite{Madrid_2019, Madrid_2019_2}, which incorporates scaled ionic charges and the rigid four-site TIP4P/2005 water model~\cite{TIP4P2005}. All interactions are described as
\begin{equation}
U_{\rm tot}^{\rm Madrid-2019} = U_{\rm el} + U_{\rm LJ},
\end{equation}
where $U_{\rm el}$ denotes the electrostatic Coulomb potential between two point charges $q^i$ and $q^j$ at a distance $r_{ij}$
\begin{equation}
\label{eq:U_charge}
U_{\rm el} = \sum_{i < j} \frac{q^i q^j}{4\pi \epsilon_0 r_{ij}},
\end{equation}
with $\epsilon_0$ the vacuum permittivity, and $U_{\rm LJ}$ describes short-range Lennard-Jones interactions  
\begin{equation}
\label{eq:U_LJ}
U_{\rm LJ} = \sum_{i < j} 4 \epsilon_{ij} \left[
\left( \frac{\sigma_{ij}}{r_{ij}} \right)^{12} - 
\left( \frac{\sigma_{ij}}{r_{ij}} \right)^6
\right].
\end{equation}
As a mean-field, effective inclusion of the electronic polarizability of the medium~\cite{Kirby2019}, the ions carry scaled charges $\pm$0.85$e$ (monovalent cations and anions) and +1.70$e$ (divalent cations) with $e$ the elementary charge. The water molecules carry partial charges $q_{\rm H} = +0.5564e$, $q_{\rm O} = 0$,  $q_{\rm M} = -2q_{\rm H}$ (the virtual M-site lies along the HOH vector bisector). All the relevant Lennard-Jones interaction parameters $\epsilon_{ij}$ and $\sigma_{ij}$ were taken from Refs.~\citenum{Madrid_2019, Madrid_2019_2}. 

Similarly to our previous work~\cite{Chubak2023Jan} on the case of NaCl (whose results are reused in the present work for comparison with the other ions), $N = 1000$ water molecules $N_{\rm p}$ ion pairs were initialized in a cubic box at the equilibrium solution density $\rho(c, T)$ obtained with $NPT$ simulations at $P = 1$~bar and $T = 298.15$~K. The resulting densities are in very good agreement with the experimental ones, as discussed in Refs.~\citenum{Madrid_2019, Madrid_2019_2}. The production runs were done in the $NVT$ ensemble. All simulations were carried out with the open-source MetalWalls package~\cite{MW2,Coretti2022} on graphics processing units with electrostatic interactions computed with Ewald summation~\cite{AguadoMadden} and a short-range cutoff of 1.24~nm, using the velocity Verlet algorithm with a time step of 1~fs and a Nosé-Hoover chain thermostat with a time constant of 1~ps. Water molecules are treated as rigid using the RATTLE algorithm with a precision of $10^{-9}$. For each system, five independent runs of 5~ns were performed to sample (every 50~fs) the EFG at the ion positions. The EFG was computed on these configurations using full Ewald summations~\cite{AguadoMadden} implemented in MetalWalls~\cite{Chubak2021Oct}. 

\subsection{Sternheimer approximation}
\label{sec:Sternheimer}

We employ the Sternheimer approximation \cite{Sternheimer1950}, in which a linear relation between the total EFG tensor ${\mathbf{V}}$ at the nucleus and the external EFG ${\mathbf{V}}^{\rm ext}$ is assumed, $\mathbf{V} \simeq (1 + \gamma) \mathbf{V}_{\rm ext}$, with $\gamma$ being the Sternheimer (anti-)shielding factor \cite{Foley1954, Sternheimer1966}. Specifically, it is assumed that the electronic cloud countribution to the EFG amplifies the field gradient induced by the surrounding charge distribution, yielding values of $\gamma$ that are typically large~\cite{Foley1954, Sternheimer1966, Carof2014, Chubak2021Oct}. The quadrupolar relaxation rate in Eq. \eqref{eq:nmr_rates} can thus be rewritten as
\begin{equation}
\label{eq:nmr_rates2}
\frac{1}{T_1} = C_Q (1+\gamma)^2 \left\langle \mathbf{V}_{\rm ext}^2 \right\rangle \tau_c,
\end{equation}
with the constant $C_Q \equiv \frac{1}{20}
\frac{2I+3}{I^2(2I-1)} \left(\frac{eQ}{\hbar}\right)^2$,
the external EFG variance
$\left\langle \mathbf{V}_{\rm ext}^2 \right\rangle \equiv 
\left\langle {\mathbf{V}}_{\rm ext}(0)\,{:}{\mathbf{V}}_{\rm ext}(0) \right\rangle$, and the correlation time 
$\tau_c$
\begin{equation}
\label{eq:tauc}
\tau_c = \left\langle \mathbf{V}_{\rm ext}^2 \right\rangle^{-1} \,
\int_0^{\infty} {\rm d}t \,
\left\langle {\mathbf{V}}_{\rm ext}(t)\,{:}{\mathbf{V}}_{\rm ext}(0) \right\rangle
\; .
\end{equation}

The electric field gradient $\mathbf{V}_{\rm ext}$ arising from the external charge distribution surrounding an ion is generally unknown. Here, we approximate it using the point charge distribution induced by the employed classical force field~\cite{Carof2014, Chubak2021Oct}. To relate this force-field-based $\mathbf{V}_{\rm ext}$ to the total EFG, $\mathbf{V}_{\rm AI}$, which can be obtained from ab initio calculations, we introduce the effective Sternheimer factor $\gamma_{\rm eff}$. This factor is determined for a given nucleus by correlating $\mathbf{V}_{\rm AI}$ with $\mathbf{V}_{\rm ext}$ on the same set of configurations of the aqueous solution~\cite{Carof2014, Chubak2021Oct}. The Sternheimer factors at infinite dilution $\gamma_{\rm eff}$ were previously calculated for Li$^+$, Na$^+$, K$^+$, Cl$^-$, Mg$^{2+}$, and Ca$^{2+}$ ions by comparing the classical $\mathbf{V}_{\rm ext}$ and ab initio $\mathbf{V}_{\rm AI}$ values of the EFG at the ion position, using configurations generated with the Madrid-2019 force field~\cite{Madrid_2019, Madrid_2019_2}, as detailed in Ref.~\citenum{Chubak2021Oct}. 

At that time, however, the Rb$^{+}$ and Cs$^{+}$ ions were not yet included in the Madrid-2019 force field, so the corresponding effective Sternheimer factors could not be determined. In 2022, the Madrid-2019 force field was extended \cite{Madrid_2019_2} to include Rb$^{+}$ and Cs$^{+}$, allowing us to compute their Sternheimer factors for this force field.  Thus, to evaluate the validity of the Sternheimer approximation and determine the corresponding $\gamma_{\rm eff}$ for Rb$^{+}$ and Cs$^{+}$, we followed the same methodology as in Ref.~\citenum{Chubak2021Oct} by comparing the cartesian components of the EFG tensor obtained from classical molecular dynamics simulations, $\mathbf{V}_{\alpha\beta}^{\mathrm{ext}}$, with those obtained from electronic DFT calculations, $\mathbf{V}_{\alpha\beta}^{\mathrm{AI}}$. For each ion, we performed five independent classical MD simulations, using the Madrid-2019 force field, of a small system consisting of 64 water molecules and a single ion. The initial configurations were prepared following the same procedure as mentioned above for the larger systems. For the five independent systems, we sampled 1000 configurations every 10~ps during a single $NVT$ production run. These configurations were then used as input for the DFT-based computations of the EFG in the condensed phase, which includes the electronic contributions. The ab initio calculations were performed using the Quantum Espresso (QE) package with the projector augmented wave (PAW) method \cite{Varini2013Aug}, which allows for a more accurate representation of the core region. The Perdew-Burke-Ernzerhof (PBE) functional was employed, with a kinetic energy cutoff of 80 Ry. Additionally, it is important to note that the GIPAW pseudopotentials \cite{Tantardini2022May} commonly used for NMR investigations were not available for Rb$^{+}$ and Cs$^{+}$ so that we used the KJPAW \cite{DalCorso2014Dec} pseudopotentials for these two ions.

In Ref.~\citenum{Chubak2023Jan}, we showed that for Na$^+$ the Sternheimer factor only slightly depends on the salt concentration in solution (likely due to modifications in the ionic hydration shell), and that one could obtain sufficiently accurate predictions using the value at infinite dilution. Furthermore, we observed in this case that the classical EFG tends to underestimate the ab initio value $\mathbf{V}_{\rm AI}$ by more than 20$\%$. For more accurate predictions of the EFG variance, we introduce a modified Sternheimer factor, $\gamma'_{\rm eff}$, such that $(1 + \gamma'_{\rm eff})^{2} = \langle \mathbf{V}_{\rm AI}^2 \rangle / \langle \mathbf{V}_{\rm ext}^2 \rangle$, using the values of $\langle \mathbf{V}_{\rm AI}^2 \rangle$ and $\langle \mathbf{V}_{\rm ext}^2 \rangle$, resulting in an EFG variance prediction within 5$\%$ accuracy. Thus, for all the ions considered in the present work, we estimated  $\gamma'_{\rm eff}$ using the ratio $\langle \mathbf{V}_{\rm AI}^2 \rangle / \langle \mathbf{V}_{\rm ext}^2 \rangle$ from the data available in Ref.~\citenum{Chubak2021Oct}, except for Rb$^+$ and Cs$^+$, for which we used the new data obtained here. The Sternheimer factors $\gamma _{\rm eff}$ and $\gamma'_{\rm eff}$ for all ions are presented in Section~\ref{sec:Sternheimer}.

\subsection{Dynamical properties}
\label{sec:Dynamics}

The EFG autocorrelation function $C_{\rm EFG}(t)$ is computed from the components $V_{\alpha\beta}(t)$ sampled every 50~fs of the larger systems described in Section~\ref{sec:MD}. In order to link the quadrupolar NMR relaxation with the dynamics of the electrolyte solution, for each salt and concentration, we consider the viscosity and ion diffusion coefficients. The shear viscosity is computed from the Green-Kubo relation~\cite{Ting2009}:
\begin{equation}
\label{eq:visc}
\eta = \frac{V}{k_{\rm B}T} \int_{0}^{+\infty} {\rm d}t \, C_{\rm stress}(t),
\end{equation}
with $V$ the volume, $k_{\rm B}$ the Boltzmann constant, and $C_{\rm stress}(t)$ the ACF of the symmetrized and traceless stress tensor $P_{\alpha\beta}$ \cite{Ting2009}:
\begin{equation}
\label{eq:stressacfs}
 C_{\rm stress}(t) = \frac{1}{10} \, \sum_{\alpha,\beta} \langle P_{\alpha\beta}(t) P_{\alpha\beta}(0) \rangle,   
\end{equation}
where $\alpha, \beta$ run over the three Cartesian components and 
$P_{\alpha\beta} = \frac{1}{2} \left( \sigma_{\alpha\beta} + \sigma_{\beta\alpha} \right) -
\frac{1}{3} \delta_{\alpha\beta} \sum_{\gamma} \sigma_{\gamma\gamma}$. The stress tensor $\sigma_{\alpha\beta}$ is sampled every step of the MD trajectory (1~fs). The diffusion coefficients are obtained from the long time limit of the mean-square displacements:
\begin{equation}
D = \lim_{t \rightarrow \infty} \frac{1}{6 M t} \, 
\sum_{i = 1}^{M} 
\left\langle \left[ \mathbf{r}_i(t) -  \mathbf{r}_i(0) \right]^2 \right\rangle,
\label{eq:Dif_coeff}
\end{equation}
where $M$ is the number of ions or water molecules, $\mathbf{r}_i(t)$ is the $i$-th particle position at time $t$, and the brackets $\langle \cdots \rangle$ denote ensemble averaging. We apply the Yeh-Hummer relation \cite{YehHummer} to account for the
finite simulation box size:
\begin{equation}
\label{eq:YehHummer}
D_{\infty} = D + \frac{k_{\rm B}T \xi}{6 \pi \eta L}
\end{equation}
with $D_{\infty}$ corresponding to the diffusion coefficient in a macroscopic system, 
$D$ that in a cubic simulation box with side length $L$, and $\xi \approx 2.837297$. 
The finite-size correction amounted to around 20\% of the measured value $D$.

Within the Stokes-Einstein-Debye (SED) model \cite{debye1929polar,einstein1956}, the time scale of EFG fluctuations is obtained as:
\begin{equation}
 \tau _{c}^{{\rm SED}} = \frac{4 \pi \eta r_{0}^{3}}{3k_{B}T},
 \label{eq:tauSED}
\end{equation}
where $\eta$ is the dynamics viscosity of the medium and $r_{0}$ is the ion’s hydrodynamic radius obtained using the Stokes-Einstein relation:
\begin{equation}
 r_{0} = \frac{k_{B}T}{6 \pi \eta D}.
 \label{eq:rzero}
\end{equation}


\section{Results and discussion}

\label{sec:RESULTS}

\subsection{Sternheimer approximation for Rb$^+$ and Cs$^+$}
\label{sec:resultsSternheimer}

\begin{figure}[ht!]
    \centering
    \includegraphics[width=\textwidth]{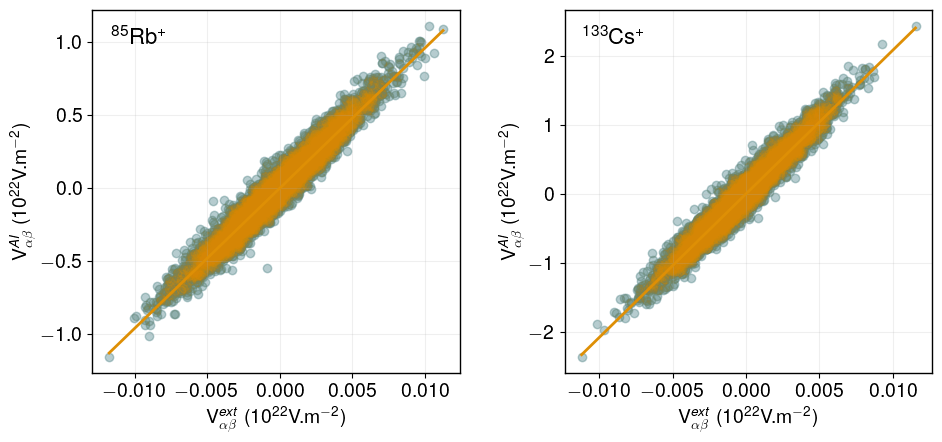}
    \caption{Comparison between the ab initio EFG components, V$^{\rm AI}_{\alpha \beta}$, (computed using the QE-KJPAW package) and the classical ones from the Madrid-2019 force field, V$^{\rm ext}_{\alpha \beta}$, at the position of Rb$^{+}$ and Cs$^{+}$ ions at infinite dilution in water. The colors correspond to 2 sets of 1000 configurations from 2 independent MD simulations. In each panel, the solid line indicates the linear fit $V^{\rm AI}_{\alpha \beta}$ = (1 + $\gamma_{\rm eff}$) $V^{\rm ext}_{\alpha \beta}$.}
    \label{fig:SternheimerRbCs}
\end{figure}

Fig.~\ref{fig:SternheimerRbCs} shows the components of the EFG tensor computed from DFT calculations, $V^{\rm AI}_{\alpha \beta}$, as a function of the ones computed with the Madrid-2019 force field, $V^{\rm ext}_{\alpha \beta}$, for aqueous Rb$^+$ and Cs$^+$ ions. In each case, results are shown only for 2 of the 5 sets of 1000 configurations for clarity. The correlation between the classical and ab initio results is comparable to that obtained with other ions (see Ref.~\citenum{Chubak2021Oct}). It confirms the relevance of the Sternheimer approximation for these ions, and that the variance $\langle \mathbf{V}_{\rm AI}^{2} \rangle$ is underestimated by $(1 + \gamma_{\mathrm{eff}})^2 \langle \mathbf{V}_{\rm ext}^{2} \rangle$, due to the spread around the linear correlation. As explained in Section~\ref{sec:Sternheimer}, we therefore define a modified Sternheimer factor as $(1 + \gamma'_{\mathrm{eff}})^{2} = \langle \mathbf{V}_{\rm AI}^{2} \rangle / \langle \mathbf{V}_{\rm ext}^{2} \rangle$. Since this was not previously done, we also computed the same quantity for the other ions using the data available in Ref.~\citenum{Chubak2021Oct}. All the Sternheimer and modified Sternheimer factors corresponding to the Madrid-2019 force field are summarized in Table~\ref{tab:SternResults}.

\begin{table}[ht!]
\centering
\begin{tabular}{|c|c|c|}
\hline
\textbf{Ion} & \textbf{$\gamma_{\rm eff}$} & \textbf{$\gamma'_{\rm eff}$} \\
\hline
Li$^+$  & 0.362 $\pm$ 0.01              & 0.42 $\pm$ 0.02 \\
Na$^+$  & 10.54 $\pm$ 0.11               & 12.09  $\pm$ 0.14  \\
K$^+$   & 28.51 $\pm$ 0.33              & 34.57 $\pm$ 0.43 \\
Rb$^+$  & 94.57 $\pm$ 0.34     & 97.35 $\pm$ 0.36 \\
Cs$^+$  & 207.05 $\pm$ 0.76    & 213.71 $\pm$ 0.76 \\
\hline
Mg$^{2+}$ & 18.34 $\pm$ 0.17            & 20.56 $\pm$ 0.33 \\
Ca$^{2+}$ & 36.24 $\pm$ 0.30            & 39.54 $\pm$ 0.59\\
\hline
Cl$^-$  & 20.25 $\pm$ 0.22               & 23.52 $\pm$ 0.03 \\
\hline
\end{tabular}
\caption{Effective Sternheimer factors ($\gamma_{\rm eff}$) and modified Sternheimer factors ($\gamma'_{\rm eff}$), for different ions at infinite dilution in TIP4P/2005 water. The values for Li$^+$, Na$^+$, K$^+$, Cl$^-$, Mg$^{2+}$, and Ca$^{2+}$ were obtained (directly for $\gamma_{\rm eff}$, and as described in the text for $\gamma'_{\rm eff}$) from our previous studies~\cite{Chubak2021Oct, Chubak2023Jan}, while those for Rb$^+$ and Cs$^+$ were computed in this work.
\label{tab:SternResults}
}
\end{table}

The values of $\gamma'_{\rm eff}$ are slightly larger by ~12$\pm$6$\%$ than $\gamma_{\rm eff}$, consistently with the underestimation of the variance by the usual Sternheimer approximation. Both $\gamma_{\rm eff}$ and $\gamma'_{\rm eff}$ increase when increasing the number of electrons in the ion (from Li$^+$ to Cs$^+$ and from Mg$^{2+}$ to Ca$^{2+}$), reflecting the larger response of the ionic cloud to an external perturbation. We also not that for isoelectronic ions (Na$^+$ and Mg$^{2+}$; Cl$^-$, K$^+$ and Ca$^{2+}$) both $\gamma_{\rm eff}$ and $\gamma'_{\rm eff}$ increase when increasing the charge of the nucleus. 

\subsection{Relaxation of electric field gradient fluctuations}
\label{sec:EFGACF}

\begin{figure}[ht!]
    \centering
    \includegraphics[width=\textwidth]{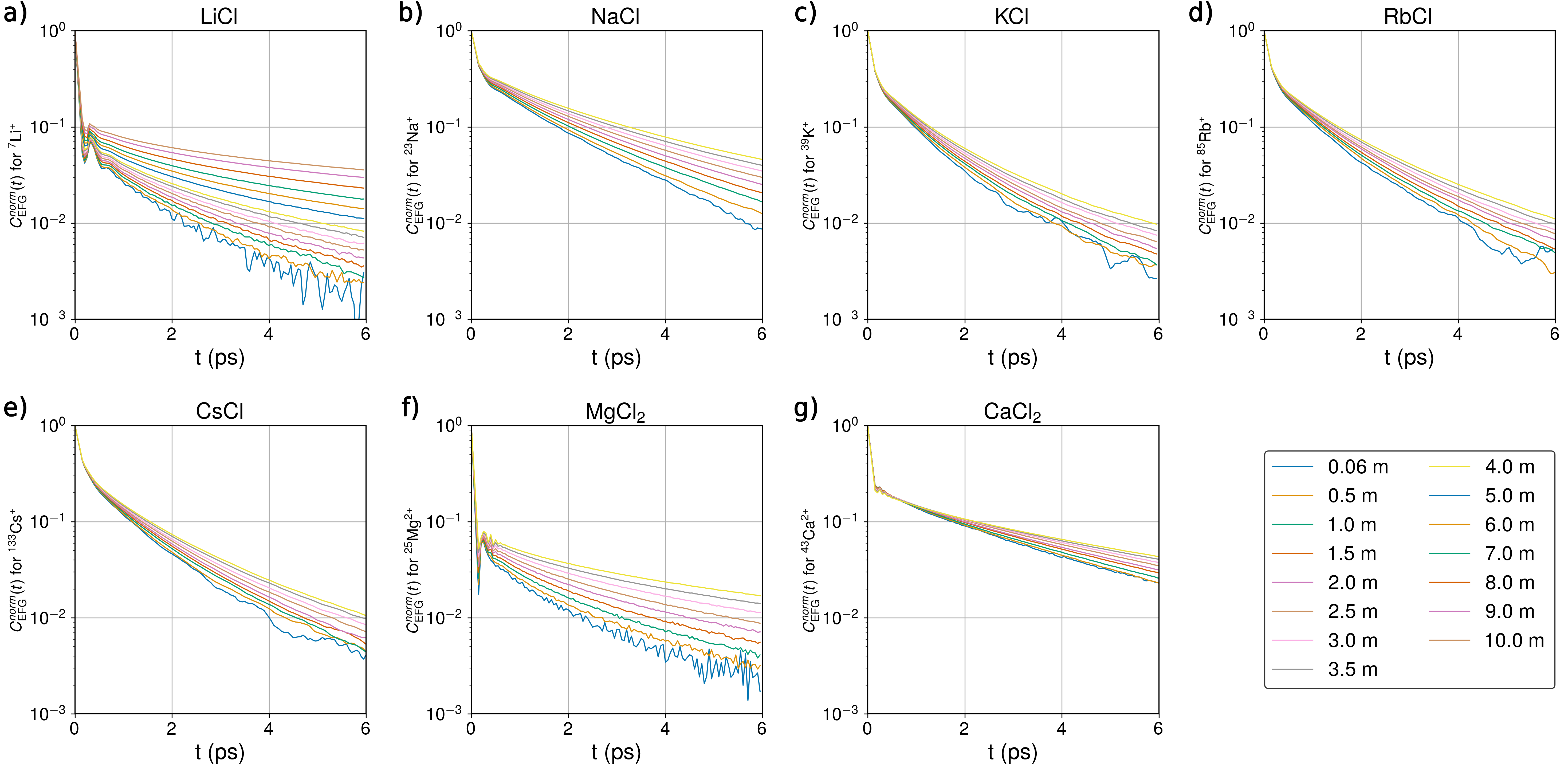}
    \caption{Normalized autocorrelation function of the electric field gradient at the cation position in aqueous solutions of (a) LiCl, (b) NaCl, (c) KCl, (d) RbCl, (e) CsCl, (f) MgCl$_{2}$, and (g) CaCl$_{2}$ as a function of time, presented for various molalities indicated by the line colors.}
    \label{fig:ACF_IONS}
\end{figure}

The normalized ACF of the classical ACF, $C_{\mathrm{EFG}}^{\mathrm{norm}}(t)$, is reported for all cations in Fig.~\ref{fig:ACF_IONS}. Each panel corresponds to a given salt and the color of the lines indicates the molality. All results exhibit similar trends with time and concentration: an initial rapid decay occurring over the first 0.2\,ps that is relatively independent of concentration, as shown more precisely in Fig.~S1 of the Supplementary Information (SI), followed by a slower decay (over several ps), which depends significantly on the salt concentration. For the lighter cations Li$^{+}$ and Mg$^{2+}$ (and to a lesser extent Ca$^{2+}$), one can also observe an intermediate oscillatory regime, due to their tight confinement within the solvation shell~\cite{Carof2015}. 

Increasing the salt concentration slows down the decay of the cation EFG ACF for all cations, as already observed for NaCl in Ref.~\citenum{Chubak2023Jan}. Such a slowdown is more pronounced for LiCl and MgCl$_{2}$. This effect is likely related to their small ionic radii~\cite{Mahler2012Jan} and the well-organized hydration shells that they form~\cite{Bernal-Uruchurtu1995Jul, Bhattacharjee2012May}. Even though the hydration shell of  CaCl$_{2}$ is similar to that of MgCl$_{2}$ in terms of hydration numbers~\cite{Friesen2019Jan}, Fig.~\ref{fig:ACF_IONS}g indicates that, within the 0 to 6~ps range, the salt concentration has a very limited effect on the slow relaxation mode of CaCl$_{2}$. However, its effect is visible on longer time scales, as shown in SI Fig.~S2, which compares $C_{\mathrm{EFG}}^{\mathrm{norm}}(t)$ at the cation positions for aqueous MgCl$_{2}$ and CaCl$_{2}$. The two systems exhibit similar behaviors, with differences arising from the larger size of Ca$^{2+}$, leading to a less structured hydration shell (also reflected in the larger variance of the EFG~\cite{Chubak2021Oct}). This analysis emphasizes the necessity of long-time sampling of the EFG to accurately determine the decorrelation time $\tau_c$, hence the relaxation rate $1/T_1$.

Similarly, for chloride ions across all seven systems, as shown in SI Figs.~S3 and~S4, $C_{\mathrm{EFG}}^{\mathrm{norm}}(t)$ displays comparable behavior, with an initial rapid decay followed by much slower secondary decays. Interestingly, the intermediate oscillatory regime observed for Li$^{+}$, Mg$^{2+}$ and Ca$^{2+}$ is barely visible for Cl$^-$ in the corresponding salt, supporting the above-mentioned interpretation as resulting from the tight solvation shell around these cations. Increasing the salt concentration induces a slowdown in the decay of $C_{\mathrm{EFG}}^{\mathrm{norm}}(t)$ at the anion positions for all systems. However, this effect is more pronounced in aqueous chloride solutions with divalent cations, MgCl$_{2}$ and CaCl$_{2}$, suggesting again the possible role of ion pairs in these systems.

In order to quantify the decay of the normalized EFG ACF of cations and anions in the seven considered electrolytes, following Ref.~\citenum{Chubak2023Jan} for Na$^+$ in aqueous NaCl we model the MD results for $C_{\mathrm{EFG}}^{\mathrm{norm}}(t)$ by a fit of the form:
\begin{equation}
f(t) = (1 - \alpha_{s})e^{-t/\tau_{f}} + \alpha_{s} e^{-[t/\tau_{s}]^{\beta}}
\label{eq:strech_fit_EFG}
\end{equation}
where the first exponential term describes the fast initial decay, with characteristic time $\tau_{f}$ never exceeding 0.12~ps, and the second stretched exponential describes the slower decay with relative weight $\alpha_s$. In practice, we first obtain $\tau_f$ for each case assuming $\beta=1$ and fitting the data to Eq.~\ref{eq:strech_fit_EFG} only for $t<1$~ps. We then fix the value of $\tau_f$ for the rest of the procedure. For each salt, we then fit the three remaining parameters ($\beta, \tau_s, \alpha_s$) using the MD results only for $t>1$~ps, for a molality $0.5$~m. The optimal values of $\beta $ for cations (resp. anions) across all seven electrolytes yielded a mean value of $ \beta \approx 0.59 \pm 0.11 $ (resp. $ \beta \approx 0.65 \pm 0.11 $), which suggests a broad distribution of relaxation modes. The values for each salt are provided in SI Table~II. The values of $\beta$ obtained at $0.5$~m are then used for all other molalities of the same salt, and only $\tau_s$ and $\alpha_s$ are fitted. The fits to Eq.~\ref{eq:strech_fit_EFG}, shown for all salts in SI Figures~S5 and~S6, describe the MD results very well. Moreover, by integrating Eq.~\ref{eq:strech_fit_EFG} for all the cations and anions (see Figs.~S7 and~S8), we find that for all systems, the slow decay provides the main contribution to the overall EFG decorrelation time $\tau_c$ obtained by integrating $C_{\mathrm{EFG}}^{\mathrm{norm}}(t)$, suggesting a broad distribution of relaxation modes in the overall EFG relaxation.

\subsection{Quadrupolar NMR relaxation rates}
\label{sec:relaxationrates}

\begin{figure}[ht!]
    \centering
    \includegraphics[width=\textwidth]{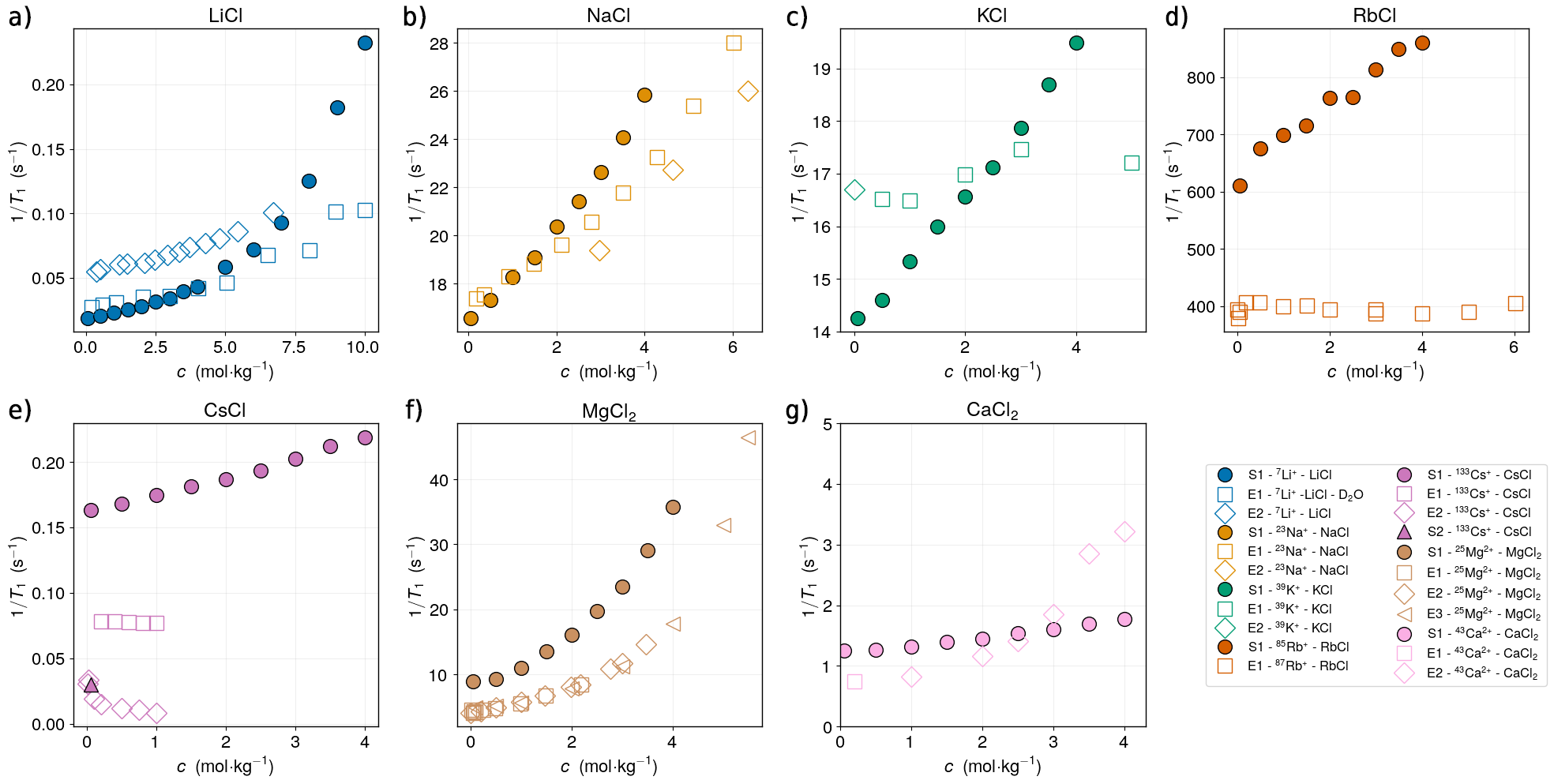}
    \caption{NMR relaxation rates as a function of the molality for six different ions $^7$Li$^{+}$, $^{23}$Na$^{+}$, $^{39}$K$^{+}$,  $^{85}$Rb$^{+}$  ($^{87}$Rb$^{+}$ for experimental results),  $^{133}$Cs$^{+}$,  $^{25}$Mg$^{2+}$ and $^{43}$Ca$^{2+}$ in seven different aqueous systems: (a) LiCl, (b) NaCl, (c) KCl, (d) RbCl, (e) CsCl, (f) MgCl$_{2}$, and (g) CaCl$_{2}$. The values are compared to experimental values, with the following references: for Li \cite{Mohammadi2020Nov,  Woessner1968Jul}, Na \cite{Chizhik1997Mar, Chubak2023Jan}, K \cite{Sahm1974Dec, fumino1997concentration}, Rb \cite{hertz1989cation}, Cs \cite{fumino1997concentration, Detscher1995May, Philips2020Sep}, Ca \cite{Fumino2002Jul,asakura1989dissolution}, Mg \cite{struis1989magnesium, Detscher1995May, Holz1979Aug}.
    }
    \label{fig:NMR}
\end{figure}

Figure~\ref{fig:NMR} compares the quadrupolar NMR relaxation rates computed by Eq.~\eqref{eq:nmr_rates2} as described above (full symbols) with that measured experimentally (open symbols). All the corresponding simulation data is available in SI Table~III. As shown in SI Fig.~S9, the convergence of the integral to compute the EFG correlation time $\tau_c$ (see Eq.~\ref{eq:tauc}) necessitates a good estimate of the normalized EFG ACF over times longer than 30~ps for the slowest decaying cases, in particular at high concentrations. This requires sampling long trajectories (several ns), which cannot be achieved with ab initio MD. However, the combination of classical MD with the effect of the electron cloud contribution via DFT calculations and the modified Sternheimer approximation provides a good estimate with experimental relaxation rates.

The NMR relaxation rates $1/T_1$ obtained from our simulations (S1 in Fig.~\ref{fig:NMR}) are of the same order of magnitude as the corresponding experimental results (E1 and E2 in Fig.~\ref{fig:NMR}) for all cations. Additionally, at a molality of 2~m, we find the following order for the relaxation rates $1/T_1$ across different cations: $^7$Li$^+$ (0.028~s$^{-1}$) $<$ $^{133}$Cs$^{+}$ (0.1868~s$^{-1}$) $<$ $^{43}$Ca$^{2+}$ (1.4480~s$^{-1}$) $<$ $^{25}$Mg$^{2+}$ (16.11~s$^{-1}$) $<$ $^{39}$K$^{+}$ (16.57~s$^{-1}$) $<$ $^{23}$Na$^{+}$ (20.38~s$^{-1}$) $<$ $^{87}$Rb$^{+}$ (763.96~s$^{-1}$), which matches perfectly the order of the experimental values available at 2m: $^7$Li$^+$ (0.062~s$^{-1}$) $<$ $^{43}$Ca$^{2+}$ (1.16~s$^{-1}$) $<$ $^{25}$Mg$^{2+}$ (8~s$^{-1}$) $<$ $^{39}$K$^{+}$ (17~s$^{-1}$) $<$ $^{23}$Na$^{+}$ (19.63~s$^{-1}$) $<$ $^{87}$Rb$^{+}$ (393.9~s$^{-1}$). Similarly, at a higher molality of 4~m, the relaxation rates $1/T_1$ for the cations from our simulations closely follow the order observed in experimental data, except for $^{25}$Mg$^{2+}$, for which we predict a $1/T_1$ larger than that of $^{23}$Na$^{+}$, while the opposite is seen in the experimental results. Overall, the NMR relaxation rates at 4~m are ranked as follows for the simulation data: $^7$Li$^+$ (0.0433~s$^{-1}$) $<$ $^{133}$Cs$^{+}$ (0.2189~s$^{-1}$) $<$ $^{43}$Ca$^{2+}$ (1.7710~s$^{-1}$) $<$ $^{39}$K$^{+}$ (19.49~s$^{-1}$) $<$ $^{23}$Na$^{+}$ (25.85~s$^{-1}$) $<$ $^{25}$Mg$^{2+}$ (35.74~s$^{-1}$) $<$ $^{87}$Rb$^{+}$ (860.74~s$^{-1}$), and for the experimental values: $^7$Li$^+$ (0.077~s$^{-1}$) $<$ $^{43}$Ca$^{2+}$ (3.226~s$^{-1}$) $<$ $^{39}$K$^{+}$ (17.21~s$^{-1}$) $<$ $^{25}$Mg$^{2+}$ (17.8~s$^{-1}$) $<$ $^{23}$Na$^{+}$ (23~s$^{-1}$)  $<$ $^{87}$Rb$^{+}$ (387.59~s$^{-1}$). 

For the first two systems, LiCl and NaCl, the behaviour of the evolution of the NMR relaxation rates from our simulations is in very good agreement with the experimental data, particularly within the concentration ranges $0~\mathrm{m} < c < 7~\mathrm{m}$ for LiCl and $0~\mathrm{m} < c < 3.5~\mathrm{m}$ for NaCl. For LiCl, the simulation results closely match the NMR relaxation data for $^7$Li$^+$ in D$_2$O up to 6~m, with a relative error of smaller than 20$\%$. As expected, the experimental rates for $^7$Li$^+$ in H$_2$O are larger, due to the additional contribution of dipolar relaxation induced by the water hydrogen spins. Therefore, the better agreement between our predicted rate at 7~m and the experimental results in H$_2$O than in D$_2$O is fortuitous. For NaCl, deviations appear around 3.5~m, after which the simulation data slightly overestimate the experimental values. However, even at 4~m, the relative error is only $\approx13\%$. Moreover, as mentioned in our previous study on NaCl only~\cite{Chubak2023Jan}, this overestimation is likely due to limitations of the employed force field in capturing the dynamic properties of solutions at higher molality, and we expect the same origin for the other salts considered in the present work. For KCl, the simulations match the experimental values closely between $1.5~\mathrm{m} < c < 3~\mathrm{m}$, with a relative error below 10$\%$. Above 3.5~m (resp. below 1~m), the simulations tend to overestimate (resp. underestimate) $1/T_1$. Nonetheless, at 0.5~m and 4~m, the relative error is $\lesssim13\%$.

For the other alkaline chlorides, RbCl and CsCl, our simulation results are generally of the same order of magnitude as the experimental data. For CsCl, the simulations exhibit a behavior consistent with experimental observations over the small range of available data. However, for both systems the simulations tend to overestimate the NMR relaxation rates, yielding values approximately 1.6 to 2.2 times larger than the experimental results, depending on the salt concentration. In the RbCl case, this overestimation could be influenced by the differences between the isotopes $^{85}$Rb$^{+}$ and $^{87}$Rb$^{+}$, even though the limited experimental data suggest that the $1/T_2$ values for $^{85}$Rb$^{+}$ are quite close to those of $^{87}$Rb$^{+}$ (see Ref.~\citenum{Lindman1972May}). Furthermore, from the few experimental results on $^{85}$Rb$^{+}$ that we found in the literature, we did observe significant variability, with a reported $1/T_2$ value of 512~s$^{-1}$ at 4~m in Ref.~\citenum{Deverell1966Jan} and a $1/T_1$ value of 710~s$^{-1}$ at infinite dilution \cite{OReilly1969Dec}. While comparing values with such differences in salt concentration is not optimal, it may contribute to the difference between the simulations and the experimental data. On the modelling side, this difference may also arise from the KJPAW pseudopotential used for calculating the effective Sternheimer factor with Rb$^+$ and Cs$^+$, for which no GIPAW pseudopotentials were available. Comparing the results with both pseudopotentials for aqueous LiCl and NaCl resulted in larger effective Sternheimer factors with KJPAW, by a factor of 1.8 and 1.5, respectively. This order of magnitude is comparable to the overestimation predicted for the quadrupolar NMR relaxation rates for $^{87}$Rb$^{+}$ and $^{133}$Cs$^{+}$. It is also possible that for these heavier ions, relativistic effects that were not taken into account in the present work might play a role in the electronic response close to the nucleus.

For the alkaline earth cations Mg$^{2+}$ and Ca$^{2+}$ in their aqueous chloride solutions, the simulation results also agree qualitatively with the experiments. For MgCl$_2$, even though the relative error is relatively high, with the simulation results being higher than the experimental values, it does follow a similar trend with molality, keeping in the range investigated range of molality studied an overestimation by a factor of $\approx2$. For CaCl$_2$, the results align well around $c = 2.4$~m but slightly overestimate (resp. underestimate) at lower (resp. higher) concentrations, with a relative error of around 40$\%$ at the lowest and highest molalities. Considering the simplicity of the force field and the challenge in the description of the solvation shell (and corresponding charge distribution) around multivalent ions, the overall agreement between the predicted and experimental relaxation rates is satisfactory.

\begin{figure}[ht!]
    \centering
    \includegraphics[width=\textwidth]{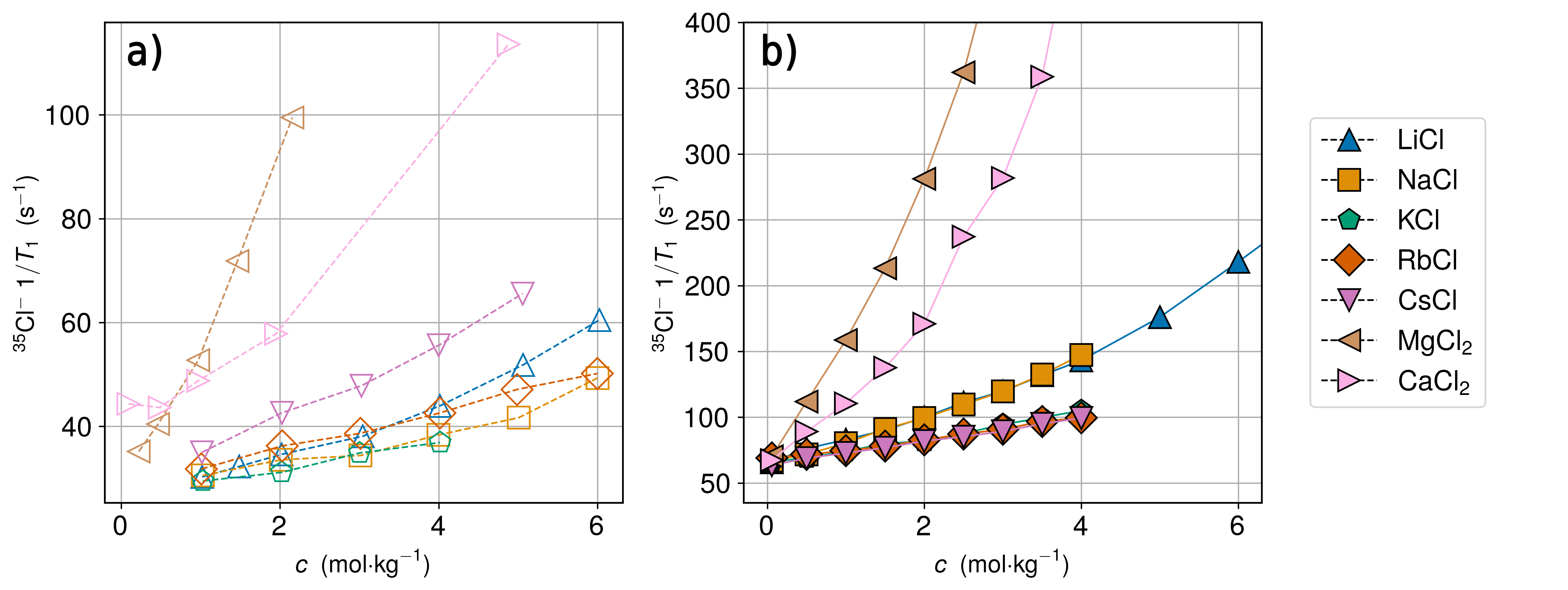}
    \caption{ 
    NMR relaxation rates as a function of the molality for $^{35}$Cl$^{-}$, in aqueous LiCl, NaCl, KCl, RbCl, CsCl, MgCl$_{2}$ and CaCl$_{2}$ solutions. Open symbols represent the experimental data in (a), whereas full symbols correspond to the predictions of the present work from our simulations in (b). The values are compared to experimental values, with the following references: LiCl \cite{Holz1977Jul}, NaCl \cite{Holz1977Jul}, KCl \cite{Holz1977Jul}, RbCl \cite{Holz1977Jul}, CsCl \cite{Holz1977Jul}, MgCl$_{2}$ \cite{struis1989magnesium} and CaCl$_{2}$ \cite{Yu2001Oct}.}
    \label{fig:NM2}
\end{figure}

Finally, Figure~\ref{fig:NM2} compares the predicted (full symbols) and experimental (open symbols) relaxation rates of the $^{35}$Cl$^{-}$ anion for all salts and molalities. As already observed for this anion at infinite dilution~\cite{Chubak2021Oct}, the Madrid-2019 force field, together with the corresponding Sternheimer factor, leads to an overestimate of $1/T_1$. Nevertheless,
the simulations reproduce qualitative trends with concentration observed experimentally. Specifically, the order of $1/T_1$ relaxation rates of Cl$^{-}$ in the different systems is CsCl $<$ RbCl $<$ KCl $<$ LiCl $<$ NaCl $<$ CaCl$_2$ $<$ MgCl$_2$ for the simulation results, whereas the experimental order is KCl $<$ RbCl $<$ NaCl $<$ LiCl $<$ CsCl $<$ CaCl$_2$ $<$ MgCl$_2$. Thus, the most notable difference is the $1/T_1$ behaviour of $^{35}$Cl$^{-}$ in aqueous CsCl; in addition, the predictions overestimates the experimental rates by a factor $\approx2-3$ in that case. For $^{35}$Cl$^{-}$ in aqueous LiCl (resp. NaCl), the overall factor is $\approx2.8$ (resp. 3.2), but is mainly due to the large overestimation (by $\approx 3.3$ and 3.8 for LiCl and NaCl, respectively) for the largest molalities where both experimental and simulation data are available. For the other electrolytes, the overestimation factors depend slightly on concentration and are $\approx 2.7$ for KCl, 2.3 for RbCl, 2.9 for MgCl$_2$ and 2.4 for CaCl$_2$.

\subsection{Static and dynamic effects on the quadrupolar NMR relaxation rate}
\label{sec:staticdynamic}

\begin{figure}[ht!]
    \centering
    \includegraphics[width=0.9\textwidth]{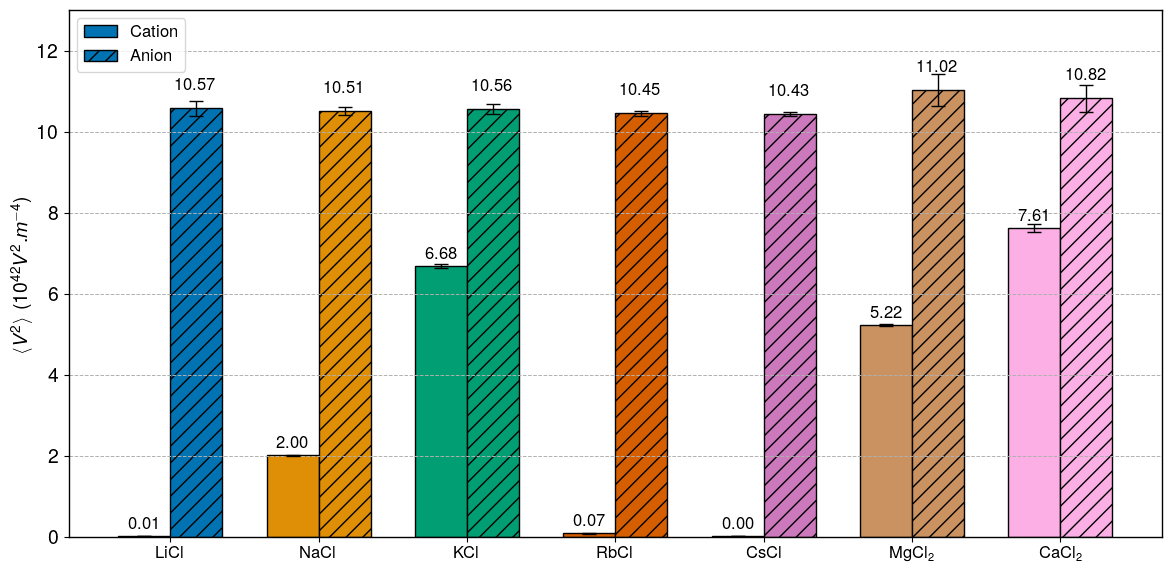}
    \caption{EFG variance $\langle V^{2} \rangle$ (including the effective Sternheimer factor) at the cation and anion positions for all the considered aqueous electrolytes. Results are shown averaged over all molalities, with errorbars indicating the associated standard deviations.
    }
    \label{fig:EFG_variance}
\end{figure}

\begin{figure}[ht!]
    \centering
    \includegraphics[width=\textwidth]{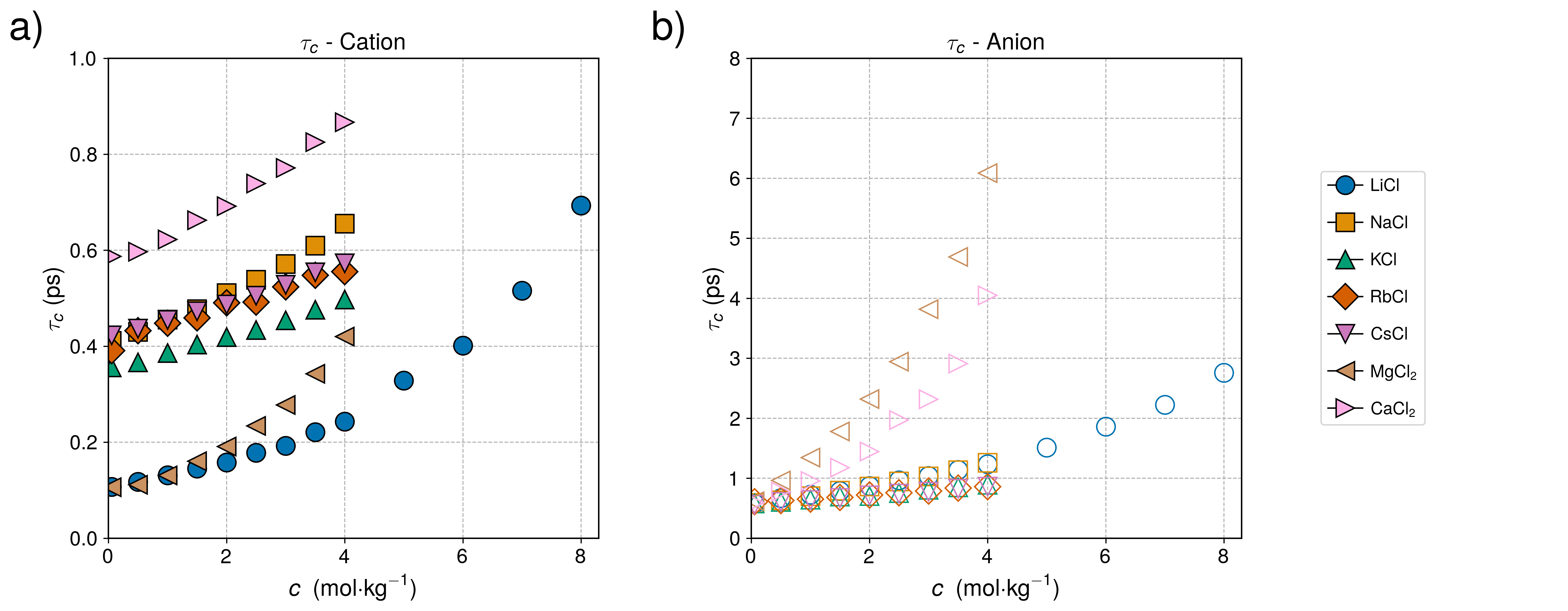}
    \caption{Effective EFG correlation time $\tau_{c}$ for the cation (a) and anion (b) in aqueous LiCl, NaCl, KCl, RbCl, CsCl, MgCl$_{2}$ and CaCl$_{2}$ solutions, as a function of molality. }
    \label{fig:EFG_correlation_time}
\end{figure}

We now investigate the role of static and dynamic effects reflected in changes in the total variance of the EFG, $\langle V^{2} \rangle$, and the effective correlation time of the EFG, $\tau_c$, at the cation and anion positions with varying salt concentrations. The results for the former, averaged over all concentrations, are presented in Fig.~\ref{fig:EFG_variance} (see SI Fig.~S10 for the effect of concentration), while the latter are shown in Fig.~\ref{fig:EFG_correlation_time}. The EFG variance for cations only slightly depends on concentration, with a maximum change of $\approx 12\%$ between 0.06~m and 4~m in aqueous MgCl$_2$, all other systems displaying variations below 4\%. In contrast, the EFG correlation time increases with concentration by factors that depend on the system. The largest increase is observed for LiCl, by a factor $\approx 11.9$ from 0.06~m to 10~m, followed by MgCl$_2$, by a factor of $\approx 9.9$ from 0.06~m to 4~m. For the other systems, this factor ranges from $\approx 1.4$ to $\approx 1.6$. 

Similarly, in the case of the anion, the total variance at the chlorine position does not change significantly with salt concentration, with the highest deviation observed in aqueous MgCl$_2$ at $\approx 12\%$ between 0.06~m and 4~m, followed by CaCl$_2$ with a deviation of $\approx 10\%$. For all other monovalent systems at the anion position, the deviation does not exceed 4$\%$. In contrast, the EFG correlation time for the three aqueous systems LiCl, MgCl$_2$, and CaCl$_2$ displays significant differences between the values at 0.06~m and at the highest molality, with factors of 7.6, 9.9, and 6.9, respectively. For all other systems, the EFG correlation time increases by a factor between 1.4 and 2.1. Overall, these observations indicate that the effect of concentration on the NMR relaxation rate for both cations and anions is primarily due to the changes in the EFG correlation time and that the effect of the changes in the variance only play a minor role. This generalizes the findings of Ref.~\citenum{Chubak2023Jan} to the whole series of alkaline and alkaline earth chloride solutions considered in the present work.

For all cations and all concentrations considered, the EFG correlation time $\tau_c$ does not exceed 1~ps (see Fig.~\ref{fig:EFG_correlation_time}a). This is consistent with our results for NaCl with the same force field~\cite{Chubak2023Jan} as well as earlier studies in the literature for various cations through classical~\cite{Engstrom1982Oct, roberts1993ionic, Carof2014, Chubak2021Oct} and ab initio MD simulations~\cite{Badu2013Sep, Philips2017Sep, Philips2020Sep}. Specifically, $\tau_c$ increases from 0.11~ps at 0.06~m to 1.27~ps at 10~m for Li$^+$, from 0.41~ps to 0.65~ps between 0.06~m and 4~m for Na$^+$, from 0.36~ps to 0.50~ps for K$^+$, from 0.39~ps to 0.56~ps for Rb$^+$, from 0.42~ps to 0.57~ps for Cs$^+$, from 0.10~ps to 0.42~ps for Mg$^{2+}$, and from 0.59~ps to 0.87~ps for Ca$^{2+}$.

From their EFG correlation time $\tau_c$ at infinite dilution, the cations can be broadly categorized into three groups: one for all the monovalent cations except Li$^+$, a second group with Li$^+$ and Mg$^{2+}$, and a third with Ca$^{2+}$ only. All cations in the first group have similar EFG decorrelation times. The second group, which corresponds to the cations with tighter hydration shells, exhibit a faster EFG decorrelation. In addition, the tetrahedral symmetry of the first hydration shell of Li$^+$ results in different EFG fluctuations compared to other ions~\cite{Carof2015}. Finally, the longest decorrelation is observed for Ca$^{2+}$, which has a less structured hydration shells. We note that the increase in $\tau_c$ with salt concentration between 0 and 4~m is more pronounced for the divalent cations, which might be related to the formation/breaking of ion pairs.

Despite these short correlation times, as mentioned above the integral defining $\tau_c$ (see Eq.~\ref{eq:tauc}) only converges over times larger by one to two orders of magnitude (see SI Fig.~S9). Importantly, for all ions the integral of the slow, concentration-dependent decay of the EFG-ACF represents the main contribution ($\approx90\%$) to $\tau_c$, and hence to $1/T_1$, as observed for NaCl and shown for all ions in Fig.~S7. This highlights the role of collective dynamics in the quadrupolar NMR relaxation of the ions, already essential at infinite dilution~\cite{Carof2015, Carof2016}, and it is slowing down as the concentration increases~\cite{Chubak2023Jan}. For the Cl$^-$ anion in all the considered aqueous salts (see Fig.~\ref{fig:EFG_correlation_time}b), one also generally finds that $\tau_c \lesssim 1$~ps, except for the divalent salts MgCl$_{2}$ and CaCl$_{2}$, as well as for LiCl beyond approximately 4~m. As for cations, $\tau_c$ increases with salt concentration and the integral of the slow, concentration-dependent decay of the EFG-ACF represents the main contribution ($\approx90\%$) to $\tau_c$, hence to $1/T_1$ (see SI Fig.~S8). 

\begin{figure}[ht!]
    \centering
    \includegraphics[width=\textwidth]{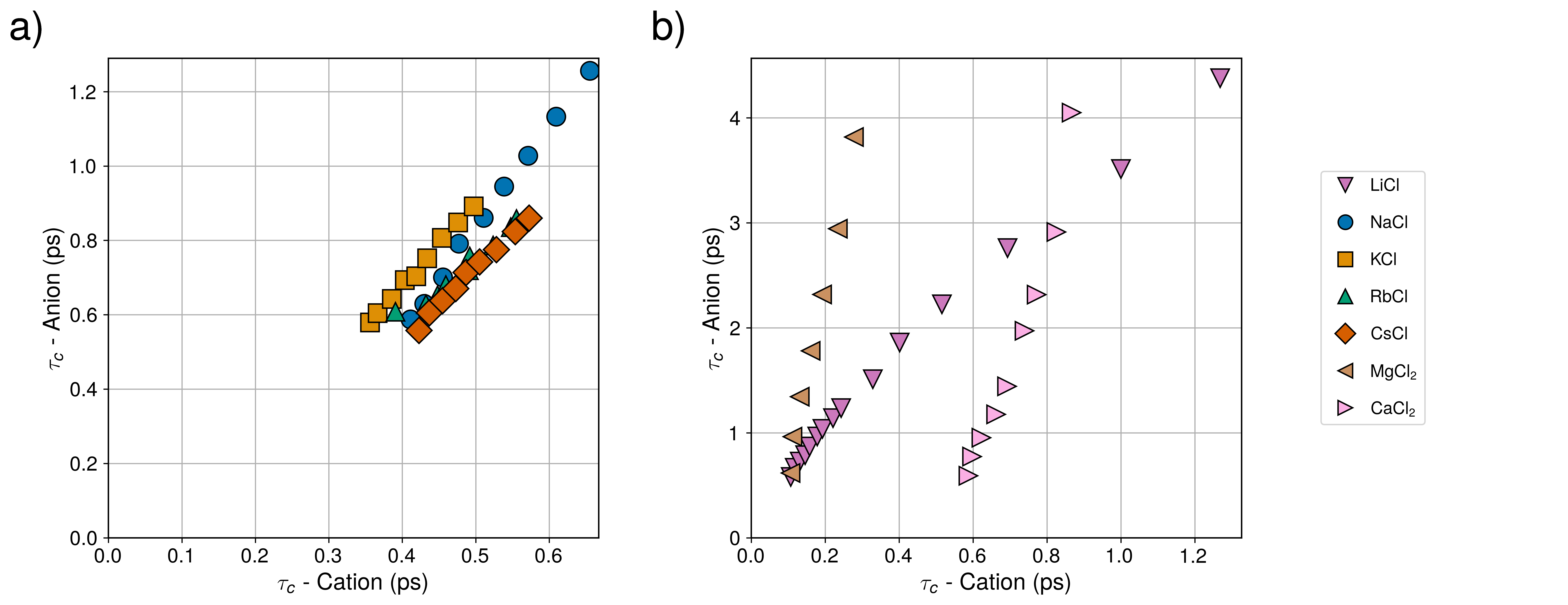}
    \caption{Correlation between the EFG correlation times $\tau_{c}$ for the cation and anion in a) aqueous NaCl, KCl, RbCl, CsCl, and b) aqueous LiCl, MgCl$_{2}$ and CaCl$_{2}$ solutions.}
    \label{fig:EFG_correlation_time_anion_cation}
\end{figure}

It is instructive to compare the EFG correlation times for the cation and anion in each system. As illustrated in Fig.~\ref{fig:EFG_correlation_time_anion_cation}a, $\tau_c$ for the cations are similar for NaCl, KCl, RbCl and CsCl, and so are $\tau_c$ for the anions in the same systems. Furthermore, the values for cations and anions are strongly correlated, with an approximately linear correlation, for this four systems the correlation time for anions being approximately 50\% larger than for cations. In contrast (see Fig.~\ref{fig:EFG_correlation_time_anion_cation}b), for LiCl, MgCl$_2$ and CaCl$_2$, even though $\tau_c$ increases for both cations and anions when increasing the salt concentration, the respective times differ markedly -- by a roughly constant factor $\approx3.5$ to 5 for LiCl, but varying from $\approx6$ to 14 for MgCl$_2$ and $\approx1$ to 5 for CaCl$_2$, depending on concentration.

Summarizing, in all systems the effect of concentration on the quadrupolar NMR relaxation rates of both cations and anions is primarily due to changes in the EFG decorrelation dynamics. Moreover, the slow decay of the EFG ACF (compared to the fast initial decay) provides the largest contribution to the overall relaxation and is well described by a stretched exponential, suggesting a broad distribution of relaxation modes arising from the coupling between the ions and the solvent dynamics. For the cations with a larger charge density (Li$^+$ with a small radius and the divalent Mg$^{2+}$ and Ca$^{2+}$), the presence of ion pairs with the Cl$^-$ may also play a role in the slowing down of the EFG fluctuations as the salt concentration increases.

\subsection{Viscosity and diffusion coefficients}
\label{sec:viscodiff}

\begin{figure}[ht!]
    \centering
    \includegraphics[width=\textwidth]{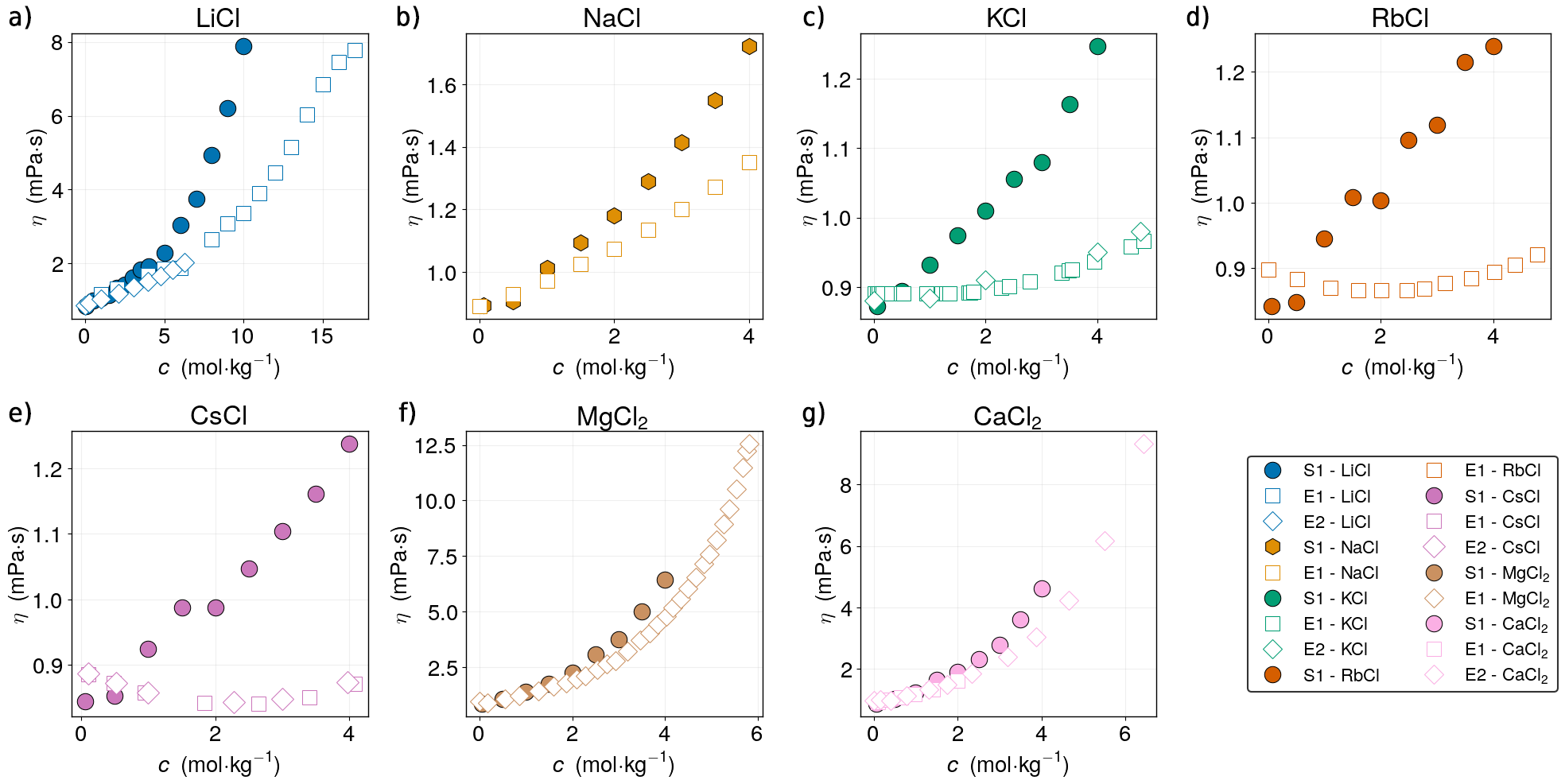}
    \caption{
    Viscosity of aqueous (a) LiCl, (b) NaCl, (c) KCl, (d) RbCl, (e) CsCl, (f) MgCl$_{2}$, and (g) CaCl$_{2}$ solutions, as a function of molality. The simulation results obtained in this work (full symbols) are compared with experimental ones (open symbols) from the literature: LiCl from~\citenum{Mohammadi2020Nov, Laliberte2007Mar, Laliberte2009Jun}, NaCl from~\citenum{Kestin1981Jan}, KCl from~\citenum{Laliberte2007Mar, Laliberte2009Jun, Zhang1996Jan}, from RbCl~\citenum{Laliberte2007Mar, Laliberte2009Jun}, CsCl from~\citenum{Nakai1995Jan}, MgCl$_{2}$ from~\citenum{Laliberte2007Mar, Laliberte2009Jun} and CaCl$_{2}$ from~\citenum{Abdulagatov2006Feb, Laliberte2007Mar, Laliberte2009Jun}.
    }
    \label{fig:visco}
\end{figure}

In order to correlate the above observations on the dynamics of the EFG fluctuations to the dynamics of the fluid, and before turning to the assessment of relaxation models in Section~\ref{sec:relaxmodels}, we first discuss the predictions of the force field for the viscosity and diffusion coefficient of ions for all electrolytes over the considered concentration range. Fig.~\ref{fig:visco} shows that the viscosity at 25$^{\circ}$C computed for all aqueous electrolytes from the stress ACF (discussed in Section~\ref{sec:relaxmodels} below) in MD simulations via Eq.~\ref{eq:visc} is overall in good agreement with experimental data from the literature, particularly at low molality. At 0.06~m, the relative error w.r.t. the experiments is below 6\% for all systems except LiCl ($\approx19$\%). The agreement deteriorates for increasing concentration, with an average overestimation factor of 1.34 $\pm$ 0.08  across all systems at 4~m. This discrepancy becomes more pronounced at high molalities, as illustrated by the aqueous LiCl system, where an overestimation factor of 2.4 is observed at 10~m. Additionally, the experimental viscosity of RbCl and CsCl solutions exhibited a weak minimum at low molality. This subtle behaviour, as reported in Ref.~\citenum{Madrid_2019_2}, is not captured by the the Madrid-2019 force field, which overestimates the viscosity for most of the concentration range for these salts.

\begin{figure}[ht!]
    \centering
    \includegraphics[width=\textwidth]{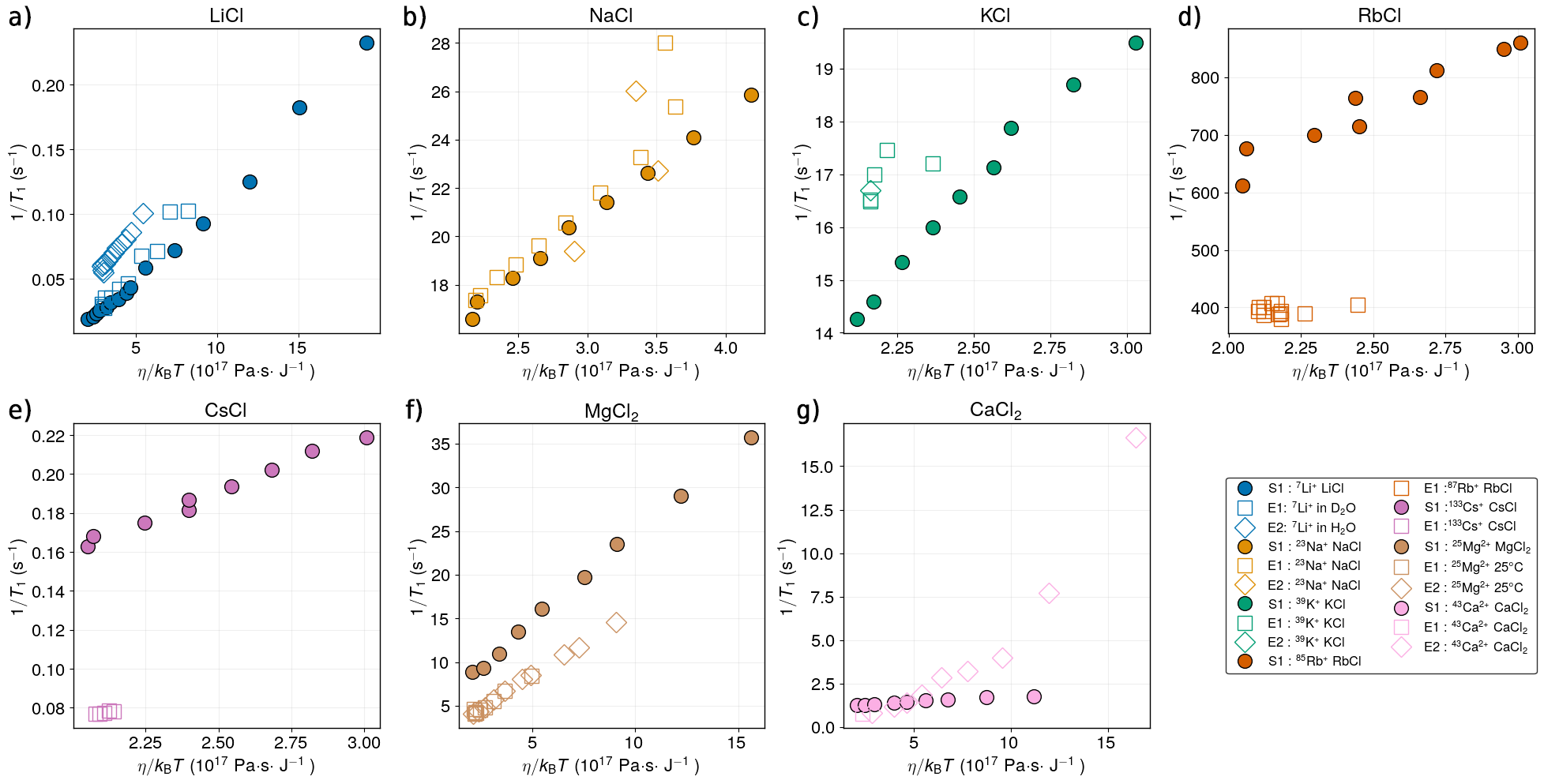}
    \caption{
    Cation NMR relaxation rates 1/T$_1$ in aqueous (a) LiCl, (b) NaCl, (c) KCl, (d) RbCl, (e) CsCl, (f) MgCl$_{2}$, and (g) CaCl$_{2}$ solutions, as a function of the ratio between viscosity and thermal energy, $\eta/k_B T$. 
    The simulation results obtained in this work (full symbols) are compared with experimental ones (open symbols) from the literature.
    }
    \label{fig:R1_kbt_cation}
\end{figure}

\begin{figure}[ht!]
    \centering
    \includegraphics[width=\textwidth]{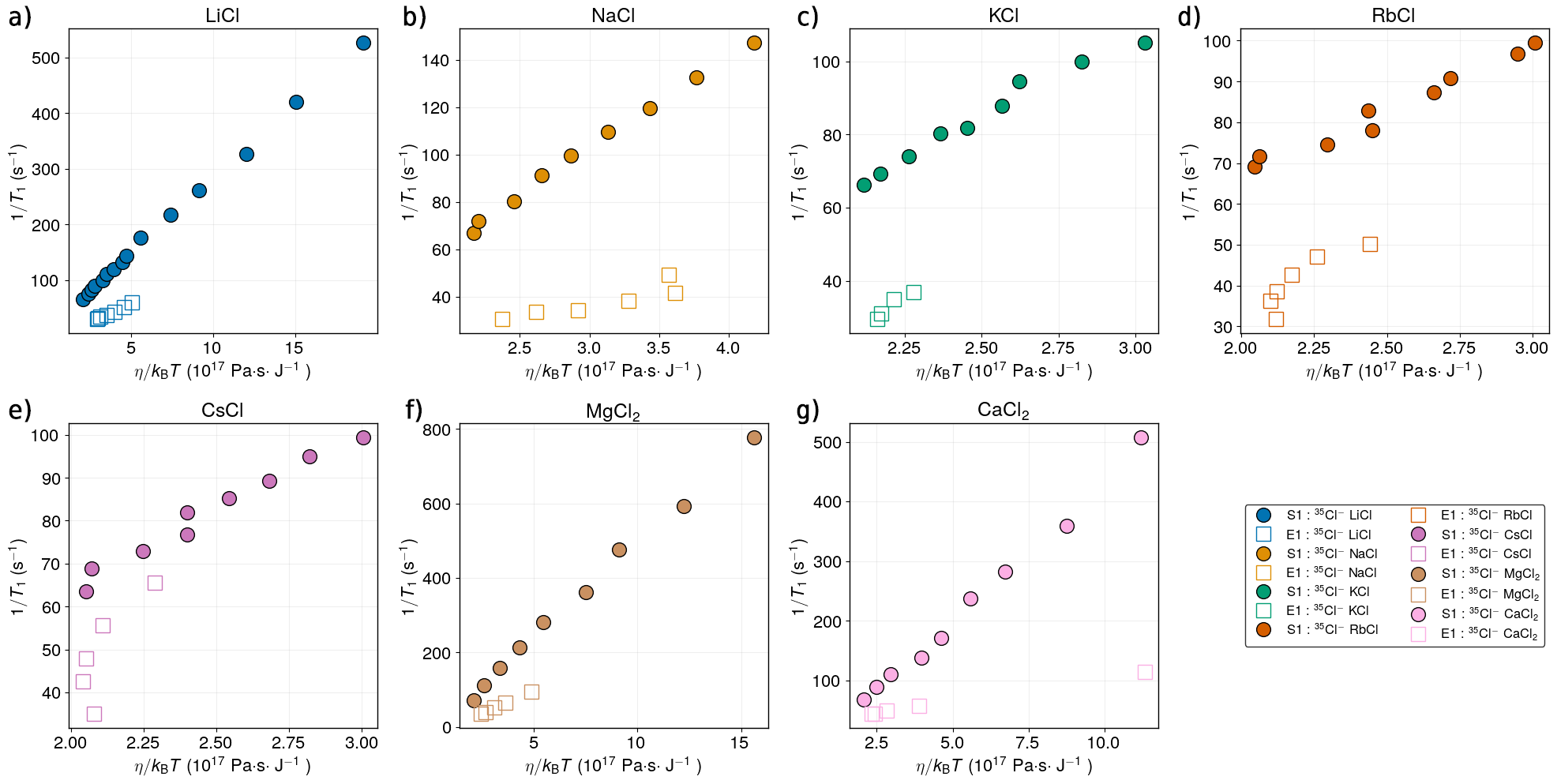}
    \caption{
    Anion NMR relaxation rates 1/T$_1$ in aqueous (a) LiCl, (b) NaCl, (c) KCl, (d) RbCl, (e) CsCl, (f) MgCl$_{2}$, and (g) CaCl$_{2}$ solutions, as a function of the ratio between viscosity and thermal energy, $\eta/k_B T$. 
    The simulation results obtained in this work (full symbols) are compared with experimental ones (open symbols) from the literature.
    }
    \label{fig:R1_kbt_anion}
\end{figure}

In previous work~\cite{Chubak2023Jan}, we have shown that the linear relation between the relaxation rate $1/T_1$ and the ratio between viscosity and thermal energy, $\eta/k_B T$, long known \emph{e.g.} for polymers or colloids, and captured by the Stokes-Einstein-Debye model, also holds for the quadrupolar relaxation of Na$^+$ cations in aqueous NaCl solutions, despite the limitations of the SED model in that case. Figs.~\ref{fig:R1_kbt_cation} and~\ref{fig:R1_kbt_anion} report $1/T_1$ as a function of $\eta/k_B T$ for cations and anions, respectively, for all considered aqueous solutions.  The simulation results obtained in this work (full symbols) are compared with experimental ones (open symbols) from the literature.

Focusing first on cations (Fig.~\ref{fig:R1_kbt_cation}), for all systems except aqueous KCl, we observe a strong correlation between $1/T_1$ and $\eta/k_B T$ for both simulations and experimental results (linear regressions lead to $R^2\geq0.9$, with a minimum for CaCl$_2$). Consistently with the results of Figs.~\ref{fig:NMR} and~\ref{fig:visco}, the agreement between simulations and experiments is excellent for aqueous LiCl and NaCl, resulting in similar slopes for the correlations between $1/T_1$ and $\eta/k_B T$: 0.01245 (resp. 4.470) for the simulations, S1, and 0.01483 (resp. 6.172) for the experiment, E1, for LiCl (resp. NaCl). Relatively good agreement is also observed for aqueous CsCl (with slopes of 0.05888, S1, and 0.02419, E1) and MgCl$_2$ (with slopes of 2.016, S1, and 1.527, E1). 

This is however not the case for RbCl (with slopes of 228.2, S1, and 15.99, E1) and CaCl$_2$ (with slopes of 0.06078, S1, and 1.100, E1), where the correlation between $1/T_1$ and $\eta/k_B T$ only separately holds for the simulations and experimental results. Several reasons might explain such discrepancy. In the RbCl case, it may derive from using the KJPAW pseudopotential to compute the effective Sternheimer factor and from the overestimation of the viscosity with the employed classical force field. For CaCl$_2$, while the simulation and experimental values of $1/T_1$ and $\eta/k_B T$ are similar, their trends with concentration are quite different. As discussed above (see Figs.~\ref{fig:NMR} and~\ref{fig:visco}), this is mainly due to the effect of concentration on $1/T_1$, which is less well predicted than on the viscosity. Finally, for KCl the simulation and experimental results are similar, even though the experiments are not well described by a linear fit.

For anions (see Fig.~\ref{fig:R1_kbt_anion}), we also observe a strong correlation between $1/T_1$ of $^{35}$Cl$^{-}$ and $\eta/k_B T$ for both simulations and experimental results. Linear regressions lead to $R^2\geq0.66$, with a minimum for CsCl, whose experimental trend of the viscosity with concentration is quite subtle, as discussed above, and all other values except for RbCl above 0.9. Overall, the correlation between $1/T_1$ and $\eta/k_B T$ for all ions suggests that ion mobility, shaped by interactions with the solvent, plays a crucial role in both the NMR relaxation rates and the viscosity of these systems. This will be further investigated in Section~\ref{sec:relaxmodels}.

\begin{figure}[ht!]
    \centering
    \includegraphics[width=\textwidth]{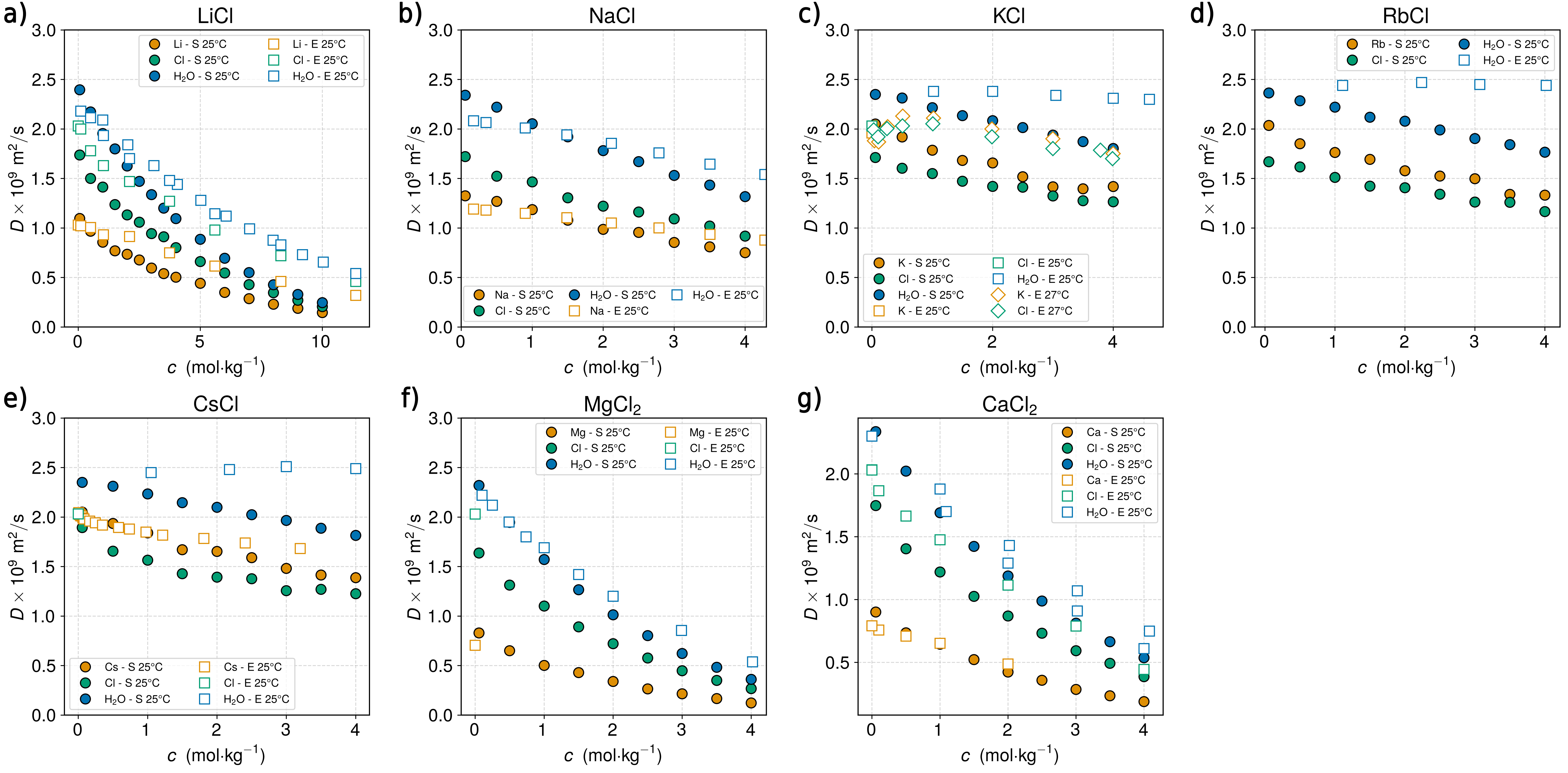}
    \caption{
    Cation (orange), anion (green) and water (blue) diffusion coefficients in aqueous (a) LiCl, (b) NaCl, (c) KCl, (d) RbCl, (e) CsCl, (f) MgCl$_{2}$, and (g) CaCl$_{2}$ solutions, as a function of molality. The simulation results (S, full symbols) are compared to experimental data (E, open symbols) from the following references: for LiCl \cite{Pethes2017Sep, Tanaka1987Jan, Muller1996Jan}, KCl \cite{Muller1996Jan, Esmaeilbeig2017Apr}, CaCl$_{2}$ \cite{Muller1996Jan, Hertz1978Apr}, MgCl$_{2}$ \cite{Muller1996Jan}, CsCl \cite{Muller1996Jan, Chakrabarti1994Feb, mills2013self}, RbCl \cite{Muller1996Jan}.
    }
    \label{fig:diff}
\end{figure}

Since the Stokes-Einstein-Debye prediction of the relaxation rate (see Eqs.~\ref{eq:tauSED} and~\ref{eq:rzero}) involves the diffusion coefficient of the ion, we compare in Fig.~\ref{fig:diff} the cation diffusion coefficient at 25$^{\circ}$C computed for all aqueous electrolytes with experimental data from the literature. While not always quantitative, especially for large salt concentrations, the simulation results are overall consistent with the experimental ones (as well as with previous simulation results with the same force field~\cite{Zeron2019Oct}, not shown).  In all cases, cation diffusion coefficients decrease with increasing concentration, and water diffuses faster than both ions. For LiCl, NaCl, MgCl$_{2}$ and CaCl$_{2}$, the anion diffuses faster than the cation, while the opposite trend is observed for KCl, RbCl and CsCl. At the lowest molalities, the relative error with respect to experiments typically ranges between $\approx$ 5-12\%; it increases to $\approx$ 20\% around 4~m and reaches $\approx$ 50-60\% in the aqueous LiCl system at 10~m. This aligns with the conclusion drawn from the viscosity that an increase in salt concentration leads to a deterioration of the predictive power of the force field for dynamical properties. Nevertheless, the agreement with experiments remains acceptable even at a few molals.

\subsection{Assessment of relaxation models}
\label{sec:relaxmodels}

Building upon the experimental and molecular simulation results presented in the previous sections, we now investigate the microscopic mechanisms underlying the NMR quadrupolar relaxation of the different cations and anions in the considered aqueous solutions. 

\subsubsection{Stokes-Einstein-Debye model}

We first consider the Stokes-Einstein–Debye (SED) model of the EFG correlation time, $\tau _{c}^{{\rm SED}}$, given by Eq.~\ref{eq:tauSED}, which corresponds to the EFG relaxation at the ion position is driven by Brownian rotational diffusion, likely linked to the collective reorientations of ion-water solvation complexes. Although the assumptions of the SED model are not expected to hold at the molecular scale~\cite{Laage2006Feb, stirnemann_mechanisms_2013, Turton2014May, laage_effect_2019, pluharova_water_2022}, we systematically explore $\tau _{c}^{{\rm SED}}$ in relation to $\tau_c$ of the different cations across the seven systems, as it is commonly used to rationalize quadrupolar relaxation dynamics in liquids.

In practice, we estimate the hydrodynamic radii $r_0$ entering in Eq.~\ref{eq:tauSED} for the various cations using the concentration-dependent diffusion coefficients and viscosities obtained from our simulations (see Eq.~\ref{eq:rzero}). This is illustrated in Fig.~S11 in the SI, which shows that $D$ is indeed proportional to $k_BT/\eta$ and provides the corresponding $r_0$. The simulated values for Li$^{+}$, Na$^{+}$, Rb$^{+}$, and Cs$^{+}$ (2.31, 1.83, 1.34 and 1.31~\AA, respectively) follow the same trend as the experimental one reported in Ref.~\citenum{Hayamizu2021Jun} (3.37, 2.44, 1.90 and 1.70~\AA, respectively), \textit{i.e.} an overestimation by a factor $\approx 1.4$. K$^{+}$ has a hydrodynamic radius (1.31~\AA) similar to Rb$^{+}$ and Cs$^{+}$, while Mg$^{2+}$ and Ca$^{2+}$ possess the largest hydrodynamic radii: 3.01~\AA\ and 2.78~\AA, respectively. The hydrodynamic radius decreases increasing ionic radius (from 0.7~\AA\ for Li$^+$ to 1.73~\AA\ for Cs$^+$, and from 0.7~\AA\ for Mg$^{2+}$ to 1.03~\AA\ for Ca$^{2+}$, see Ref.~\citenum{Marcus1988Dec}) and increasing ionic charge. These trends are consistent with the previous study of \citet{Hayamizu2021Jun}, and reflects the tighter hydration structure of ions when the electric field due to the ion is larger in the first hydration shell.

As shown in Fig.~S12 in the SI, the resulting $\tau _{c}^{{\rm SED}}$ indeed correlate with the simulation results $\tau_c$ for all cations. However, the SED model consistently overestimates the relaxation time, by a factor ranging from $\sim 6$ (for Rb$^+$ and Cs$^+$) to more than 100 for the divalent cations. Since Figs.~\ref{fig:R1_kbt_cation} and~\ref{fig:R1_kbt_anion} indicate a correlation between the relaxation rate and the ratio $\eta/k_{B}T$, as in the SED model, we extract an effective Stokes radius $r_{\rm eff}$ from the correlation between $\tau _{c}$ and $\eta/k_{B}T$, shown in Fig.~S13, using a fit of the form:
\begin{equation}
\tau_{c} = \frac{4 \pi \eta (r_{\rm eff})^{3}}{3k_{B} T} + \tau_{\rm eff},
\label{fittau}
\end{equation}
with an offset $\tau_{\rm eff}$ (absent in the standard SED model) to improve the quality of the fit. The resulting $r_{\rm eff}$) are significantly smaller than the expected ionic radii, with values ranging from 0.38~\AA\ for Mg$^{2+}$ to 0.72~\AA\ for K$^{+}$ and Cs$^{+}$. Furthermore, for some cations, the inclusion of an additional intercept $\tau_{\rm eff}$ is necessary to accurately fit the data, as highlighted by \citet{Turton2014May}, indicating the limitations of the SED model. Overall, these observations generalize  to all considered electrolytes the ones made for aqueous NaCl in Ref.~\citenum{Chubak2023Jan}, and confirm that the SED model using the hydrodynamic radius corresponding to translational diffusion cannot predict the quadrupolar relaxation time.

\subsubsection{Link between EFG and stress fluctuations}

\begin{figure}[ht!]
    \centering
    \includegraphics[width=\textwidth]{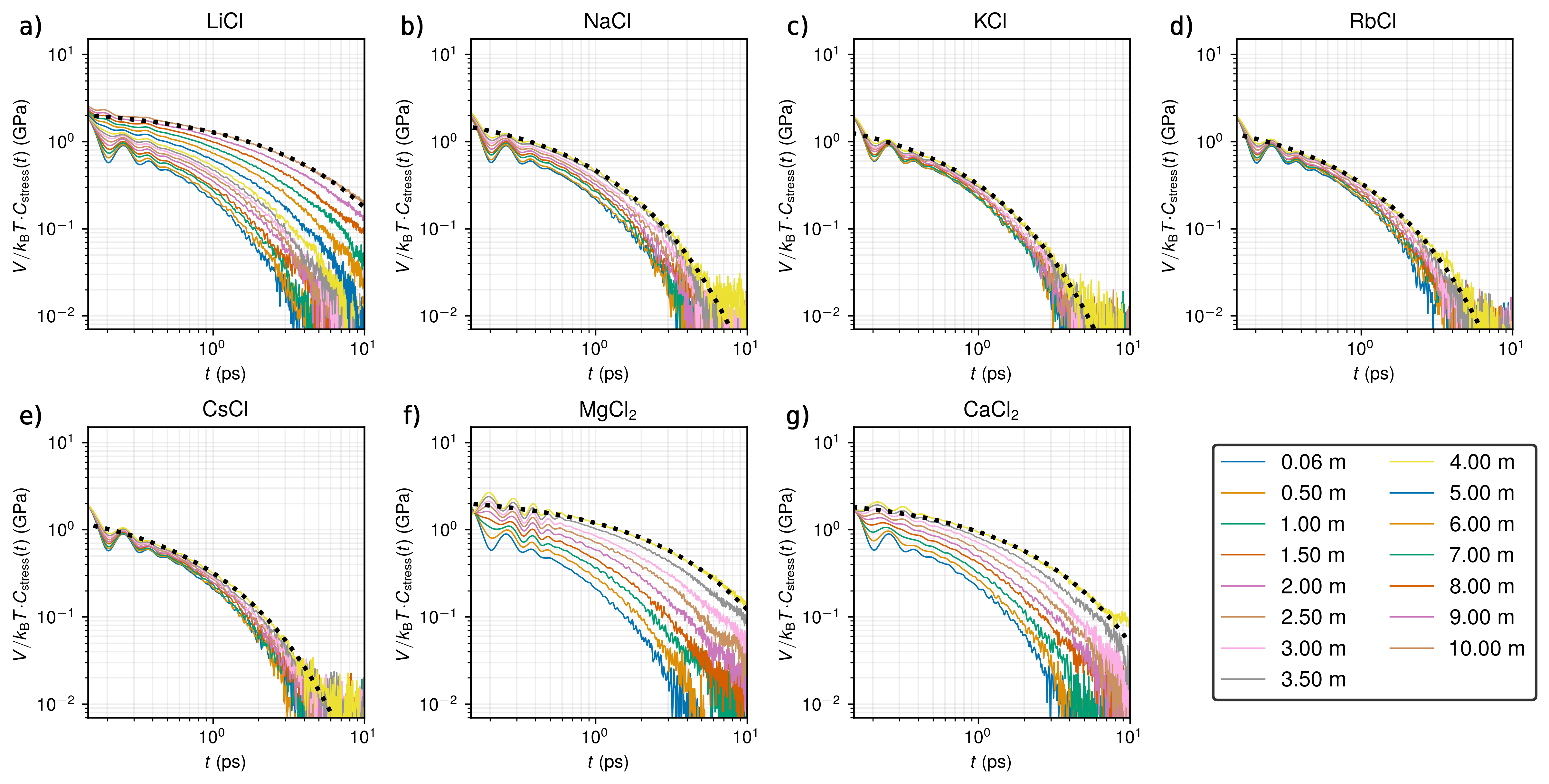}
    \caption{
     Stress autocorrelation function (see Eq.~\ref{eq:stressacfs}) of aqueous (a) LiCl, (b) NaCl, (c) KCl, (d) RbCl, (e) CsCl, (f) MgCl$_{2}$, and (g) CaCl$_{2}$ solutions, as a function of molality (indicated by the color). Results are shown multiplied by $V/k_BT$, so that their integral from 0 to $\infty$ provides the viscosity (see Eq.~\ref{eq:visc}). In each panel, the dotted black line corresponds to a fit to a stretched exponential of the decay at long time for the highest molality.}
    \label{fig:ACFCstress}
\end{figure}

For aqueous NaCl, we recently found a good correlation between the EFG relaxation time for Na$^+$ and the relaxation time of the stress tensor~\cite{Chubak2023Jan}. As the EFG, the stress is a rank-2 tensor, and the integral of its ACF is related to the viscosity (see Eqs.~\ref{eq:visc} and~\ref{eq:stressacfs}), just as that of the EFG is related to the quadrupolar NMR relaxation rate. Here we find that for all the considered electrolytes the stress ACF exhibits a long-time tail, whose decay is slowed down with increasing salt concentration, indicating an overall deceleration of the viscous dynamics of the liquid (see Fig.~\ref{fig:ACFCstress}). A characteristic timescale can be defined as usual as the integral of the normalized ACF. Since the long-time tail of $C_{\rm stress}(t)$ is well-described beyond a few tens of femtoseconds by a stretched exponential decay, ($\sim e^{-(t/\tau_{k})^{\beta_{k}}}$, shown as dashed lines in Fig.~\ref{fig:ACFCstress}), we obtain with this form the mean structural relaxation time $\tau_{\rm struct}=\tau_{k} \beta_{k}^{-1} \Gamma(\beta_{k}^{-1})$. Following Ref.~\citenum{Chubak2023Jan}, for all systems we used $\beta_k=0.61$ and considered only $\tau_k$ as a fitting parameter, with values around 0.25~ps for all systems at 0.06~m, an increase with concentration up to 0.4~ps (resp. more than 1~ps) for monovalent (resp. divalent) ions at 4~m, and a further to 2.05~ps at 10~m for aqueous LiCl. 

\begin{figure}[ht!]
    \centering
    \includegraphics[width=\textwidth]{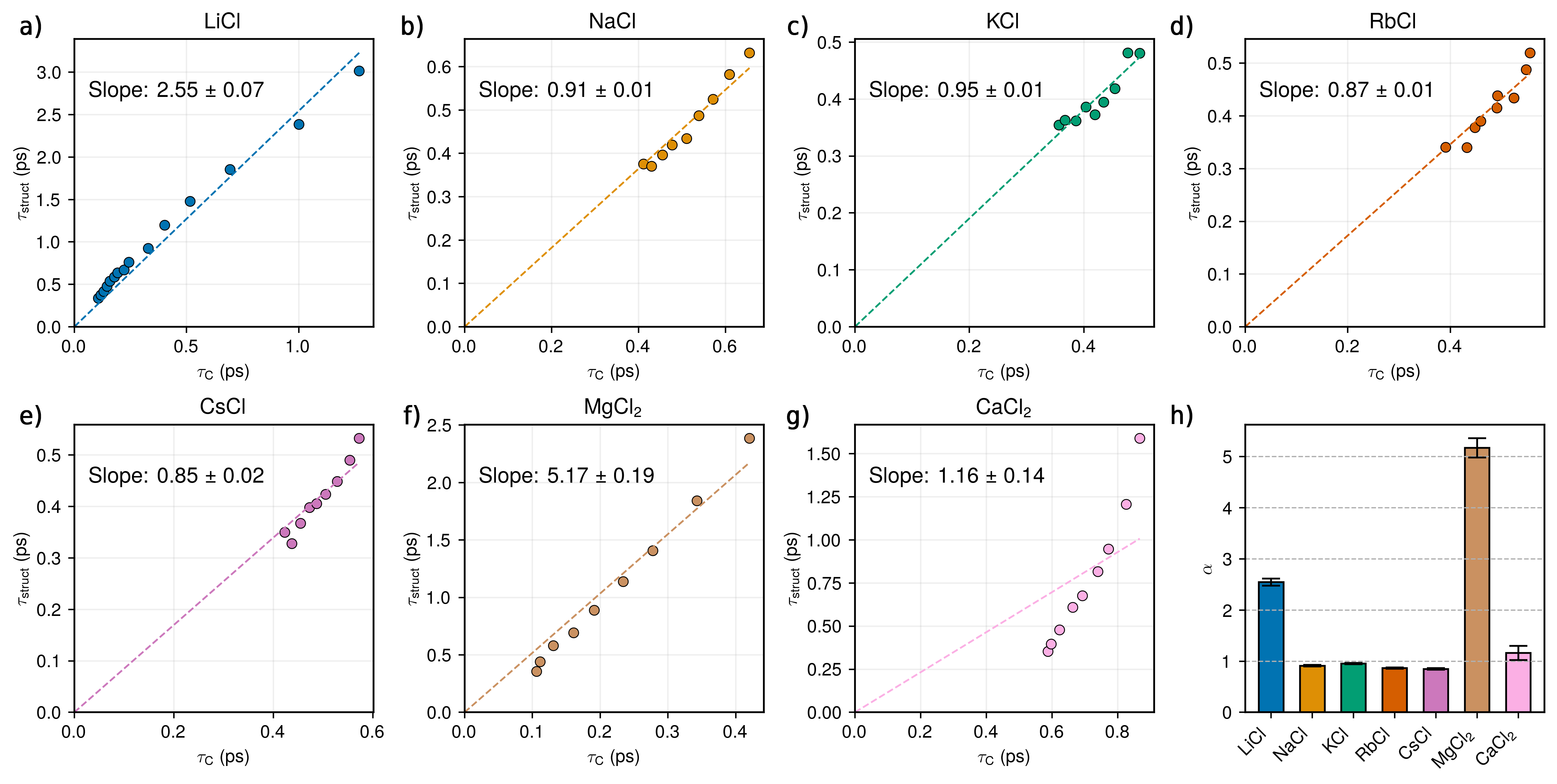}
    \caption{Comparison between the structural relaxation time $\tau_{\rm struct}$ and the EFG correlation time $\tau_{c}$ of cations in aqueous (a) LiCl, (b) NaCl, (c) KCl, (d) RbCl, (e) CsCl, (f) MgCl$_{2}$, and (g) CaCl$_{2}$ solutions. The dashed lines represent linear fits defined as $\tau_{\rm struct}$ = $a~\tau_{c}$ with the cation-specific slopes $a$ indicated in each panel and summarized in the last panel as a bar plot. The data for CaCl$_2$ is not well described by such a fit, but we provide the corresponding slope for completeness.}
    \label{fig:tau_struct_tau_c_cation}
\end{figure}

The resulting structural relaxation time $\tau_{\rm struct}$, shown for all salts as a function of concentration in Fig.~S14, is compared with the EFG relaxation time of cations, for all considered electrolytes in
Fig.~\ref{fig:tau_struct_tau_c_cation}. As in the case of NaCl in Ref.~\citenum{Chubak2023Jan}, we find in all cases except CaCl$_2$ that they are approximately proportional to each other, \textit{i.e.} $\tau_{\rm struct}\approx a~\tau_{c}$. In addition, the proportionality factor $a$ is close to one (0.91, 0.95, 0.87 and 0.85 for Na$^+$, K$^+$, Rb$^+$ and Cs$^+$, respectively) for all monovalent cations except Li$^+$ ($a\approx 2.55$). Even though the stress and EFG tensors are not directly related, both quantities are inherently collective and the decay of their ACFs is primarily governed by many-body correlations and exhibits a similar stretched decay for $t > 0.4-0.5$~ps. These observations suggest that, in all these systems, the fast collective dynamics of the liquid, which drive their structural rearrangements, play a key role in quadrupolar NMR relaxation. 

For MgCl$_2$, the linear relation still holds, but as for LiCl the proportionality factor significantly deviates from 1 ($a=5.17$). In both cases, this might be attributed to the tighter solvation shells of these ions, already mentioned when discussing the viscosity or the hydrodynamic radius, which may result in a weaker coupling with water molecules further from the ions. One can hypothesize that the collective relaxation in that case can be represented as arising from the coupling between the dynamics of (large) hydrated ions in the remaining solvent (similar to Na$^+$, K$^+$, Rb$^+$ and Cs$^+$), the internal dynamics of this species (within the hydration shell) and the exchange of solvent molecules in and out of the hydration shell. For aqueous CaCl$_2$, while there is a clear correlation between the structural relaxation times, a fit by $\tau_{\rm struct}\approx a~\tau_{c}$ is clearly not a good representation of the dat, suggesting a more complex interplay between the system's structural relaxation and the EFG relaxation of the cation.

In the specific case of LiCl, an additional deviation from the linear regime described by  $\tau_{\rm struct} = a~\tau_c$ occurs at molalities higher than 4~m, requiring an additional intercept to accurately fit the values of the two relaxation times. This deviation could arise from different sources, including i) limitations of the force field used: Deviations from experimental results are also observed at higher molalities, suggesting the force field might not fully capture the system's behaviour; ii) structural deformation: From our previous study concerning Na$^+$, $\tau_c$ appears to primarily depend on the first two hydration shells. At increasing molalities, overlapping of hydration shells is likely to occur. This could form specific hydration structures around lithium, significantly affecting $\tau_c$. The second explanation could align with previous studies \cite{Harsanyi2005Mar, Harsanyi2011Jan} suggesting that the increased molality distorts the hydration shell of Li$^+$ ions and change its hydration number.

\section{Conclusion}
\label{sec:CONCLUSIONS}

We investigated the quadrupolar NMR relaxation of cations and anions in aqueous electrolyte solutions, including NaCl, LiCl, KCl, RbCl, CsCl, MgCl$_2$, and CaCl$_2$, across a broad range of salt concentrations, using molecular simulations. To that end, we used our previouly introduced approach combining DFT-PAW calculations and classical molecular dynamics to compute the electric field gradient fluctuations over the relavant time scales. The predicted NMR relaxation rates $1/T_1$ are in good agreement with experiments from the literature, quantitative up to a few molals for some cations, and qualitative (even though overestimated) for the $^{35}$Cl$^-$ anion across the 7 aqueous solutions. 

We analyzed the contributions of static and dynamic effects to the quadrupolar NMR relaxation rates. Our results demonstrate that the growth of $1/T_1$ for the various ions with increasing salt concentration is primarily driven by a slowing down of the EFG fluctuations, and that the variance of the EFG at the ion positions does not vary significantly with concentration. For all ions, the main contribution to the EFG correlation time $\tau_c$ arises from the second, slower decay mode of the EFG ACF, which could be fitted with a stretched exponential, suggesting a broad distribution of relaxation modes. We highlighted the different behaviour of smaller (Li$^+$) and divalent (Mg$^{2+}$ and Ca$^{2+}$), compared to the other monovalent cations, which points to the effect of their tighter solvation shells and of the possible role of ion pair formation in the decay of the EFG fluctuations.

For low molalities, the experimental viscosity of all aqueous solutions is generally well reproduced by the classical force field used, although some specific behaviours, previously highlighted in Refs.~\citenum{Madrid_2019} and ~\citenum{Madrid_2019_2} are not accurately captured. The linear correlation between $1/T_1$ and the viscosity ($\eta/k_B T$), which usually motivates the use of the Stokes-Einstein-Debye model, approximately holds in these systems both in simulations and experiments. By estimating the SED relaxation time from the concentration-dependent viscosities and ion diffusion coefficients in our simulations. Even though the hydrodynamic radii deduced from the diffusion coefficients are reasonable and the SED relaxation time correlates with $\tau_c$, it is systematically larger (by at least a factor of 6-7), as observed for Na$^+$ in aqueous NaCl solution in Ref.~\citenum{Chubak2023Jan}.

In contrast, the structural relaxation time scale extracted from the fluctuations of the stress tensor introduced in the same reference is proportional to $\tau_c$ (except for aqueous CaCl$_2$), and within 15\% of its value for all monovalent salts except LiCl. Even though the stress and EFG tensors are not directly related, both quantities are inherently collective, \textit{i.e.} the relaxation of their fluctuations is primarily governed by many-body correlations. Thus, the fast collective dynamics of the liquid, which drive their structural arrangements, also play a key role in quadrupolar NMR relaxation. Similar conclusions can be drawn for LiCl and MgCl$_2$, even though their tighter solvation shells result in a faster decorrelation of the EFG fluctuations compared to that of the stress tensor. These results generalize in particular the observation for Na$^+$ in aqueous NaCl solution that models assuming a viscous model of the solvent dynamics, such as the SED, are not sufficient to describe the EFG fluctuations, since they assume a fast decay of the stress fluctuations compared to that of the EFG.

The detailed microscopic mechanisms leading to the above-mentioned structural rearrangements and their effects on both the stress and EFG relaxation remain however to be clarified, and it might be particularly useful to combine the information on the NMR relaxation of ions with that of water~\cite{zhang_range_2024} or insights from other experimental techniques. It would in particular be interesting to consider the role of discrete events whereby water molecules change of partner within the hydrogen bond network, described by the so-called "jump model"~\cite{Laage2006Feb,stirnemann_mechanisms_2013,laage_effect_2019,gomez_water_2022,pluharova_water_2022}. In addition to considering more complex systems such as water-in-salt electrolytes, which contain more ions than water, it is also relevant to develop analytical theories accounting for the effect of salt concentration on the EFG relaxation, building on implicit solvent descriptions of electrolytes such as the ones developed to predict the electrical conductivity~\cite{Illien2024}.  


\section*{Acknowledgments}
We thank Alexej Jerschow, Guillaume M\'eriguet and Anne-Laure Rollet for useful discussions. This project received funding from the European Research Council under the European Union’s Horizon 2020 research and innovation program (grant agreement no. 863473). This work was performed using HPC resources from GENCI-IDRIS (Grant-2024-AD010912966R2).

\section*{Author declarations}

\subsection*{Conflict of interest}
There is no conflict of interest to declare

\subsection*{Author contributions}
\textbf{Matthieu Wolf:} Conceptualization (equal); Formal analysis (equal); Investigation (lead); Methodology (supporting); Software (supporting); Writing/Original Draft Preparation (lead); Writing – review \& editing (equal).
\textbf{Iurii Chubak:} Conceptualization (equal); Formal analysis (equal); Investigation (supporting); Methodology (supporting); Software (lead); Writing – review \& editing (equal).
\textbf{Benjamin Rotenberg:} Conceptualization (lead); Formal analysis (equal); Funding Acquisition (lead); Investigation (supporting); Methodology (lead); Supervision (lead); Writing – review \& editing (equal).

\section*{Data availability}
\BR{To be completed consistently with the final version of the manuscript:} The original data presented in this study are openly available in Zenodo at \BR{[DOI/URL]} or \BR{[reference/accession number]}.

\bibliographystyle{rsc}
\bibliography{references}

\clearpage
\newpage

\setcounter{figure}{0}
\renewcommand{\figurename}{Fig.}
\renewcommand{\thefigure}{S\arabic{figure}}

\begin{figure}[ht!]
    \centering
    \includegraphics[width=\textwidth]{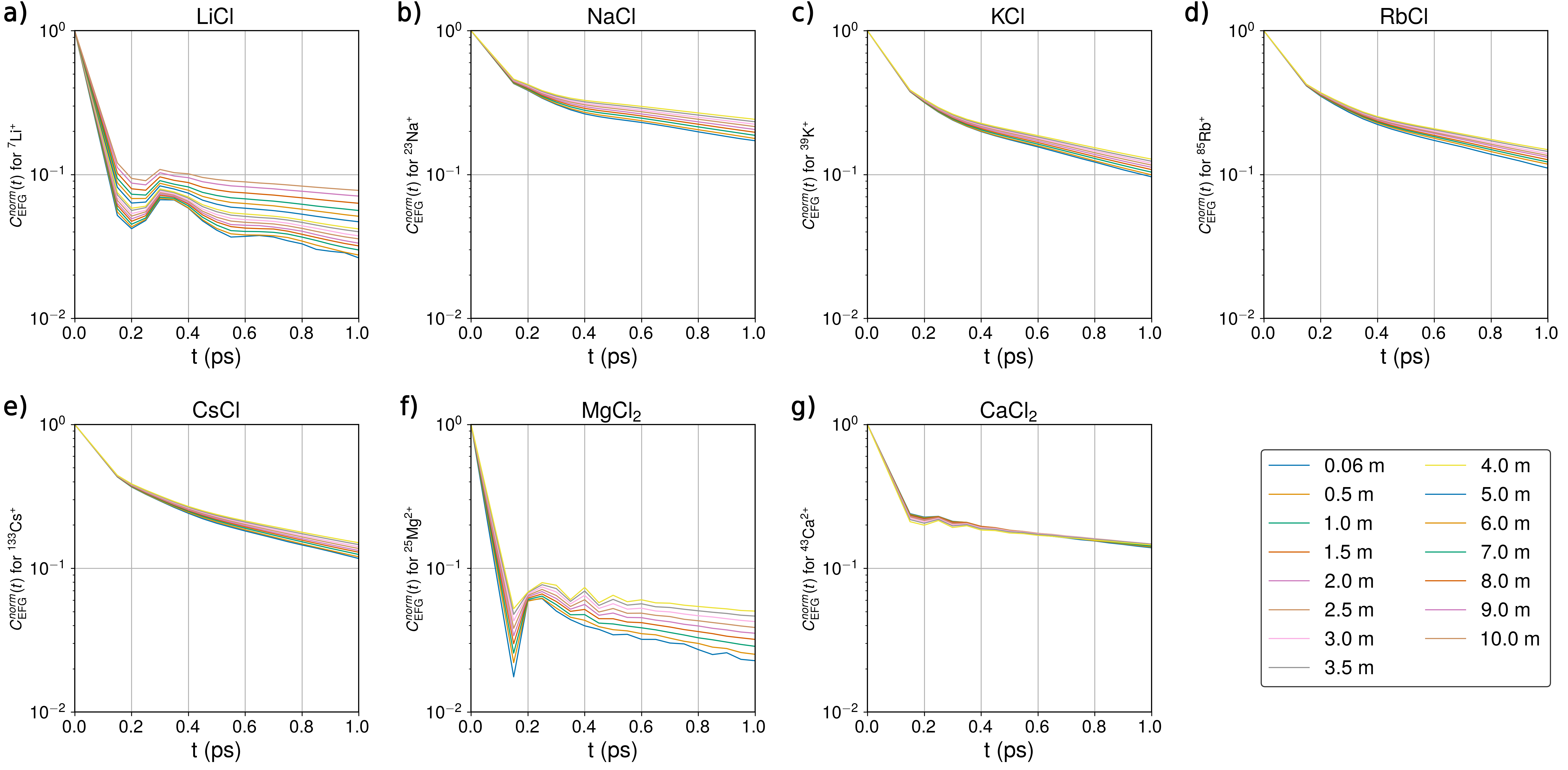}
    \caption{Normalized autocorrelation function of the electric field gradient at the cation position in aqueous (a) LiCl, (b) NaCl, (c) KCl, (d) RbCl, (e) CsCl, (f) MgCl$_{2}$, and (g) CaCl$_{2}$ as a function of time from 0 to 1~ps, presented for various molalities.}
    \label{fig:ACF_IONS_zoom}
\end{figure}

\begin{table}[ht]
\centering
\begin{tabular}{|c|c|c|}
\hline
\textbf{System} & \textbf{$\beta$ - Cation} & \textbf{$\beta$ - Anion} \\ 
\hline
\textbf{LiCl}   & 0.5202 & 0.6980 \\ 
\textbf{NaCl}   & 0.6747 & 0.6793 \\ 
\textbf{KCl}    & 0.5129 & 0.7572 \\ 
\textbf{RbCl}   & 0.6025 & 0.6813 \\ 
\textbf{CsCl}   & 0.5464 & 0.7486 \\ 
\textbf{MgCl$_2$} & 0.4583 & 0.4475 \\ 
\textbf{CaCl$_2$} & 0.8150 & 0.5095 \\ 
\hline
\end{tabular}
\caption{$\beta$ values for fits using Eq.13 for cations and anions in aqueous electrolytes. }
\label{tab:beta_values}
\end{table}

\begin{figure}[ht!]
    \centering
    \includegraphics[width=\textwidth]{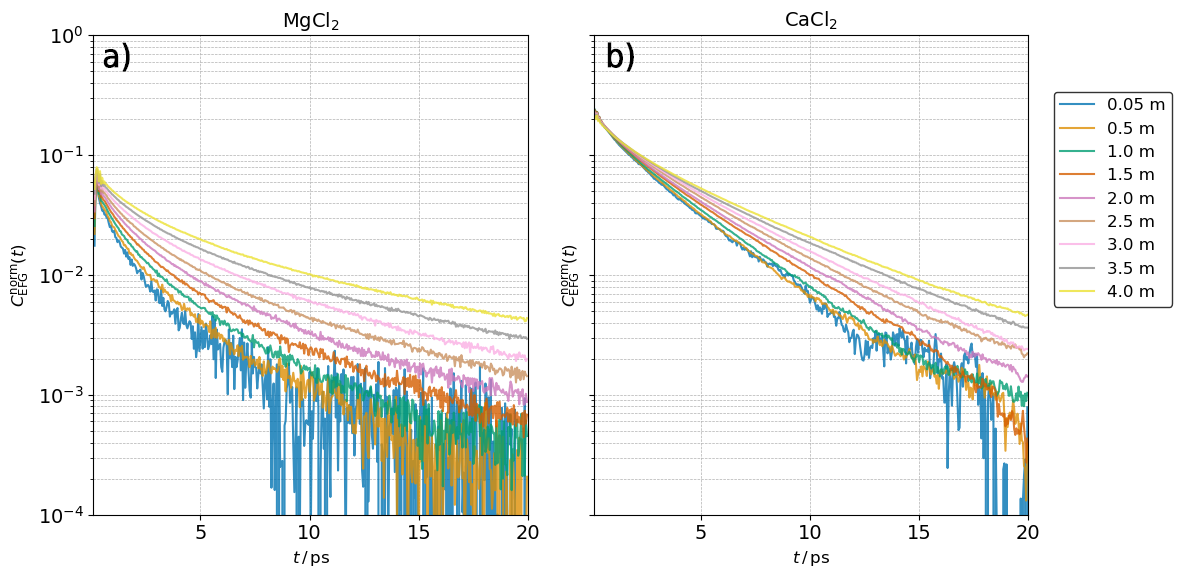}
    \caption{Normalized autocorrelation function of the electric field gradient at the cation position for MgCl$_{2}$, and CaCl$_{2}$, as a function of time from 0 to 20~ps, presented for various molalities.}
    \label{fig:ACF_IONS_MgCl2_CaCl2}
\end{figure}

\begin{figure}[ht!]
    \centering
    \includegraphics[width=\textwidth]{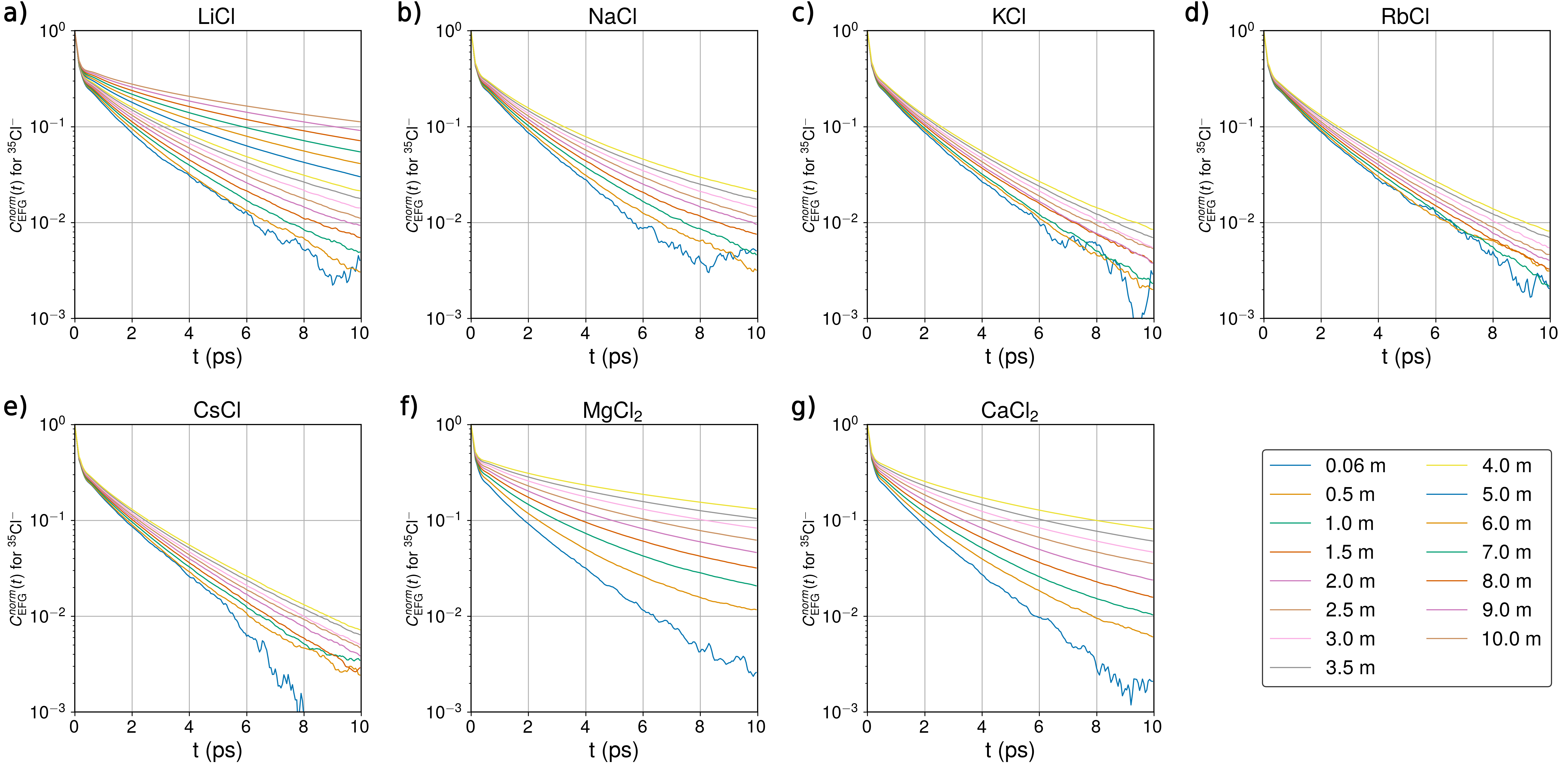}
    \caption{Normalized autocorrelation function of the electric field gradient at the anion position in aqueous solutions of (a) LiCl, (b) NaCl, (c) KCl, (d) RbCl, (e) CsCl, (f) MgCl$_{2}$, and (g) CaCl$_{2}$, as a function of time, presented for various molalities indicated by the line colors.}
    \label{fig:ACF_Chlorine}
\end{figure}

\begin{figure}[ht!]
    \centering
    \includegraphics[width=\textwidth]{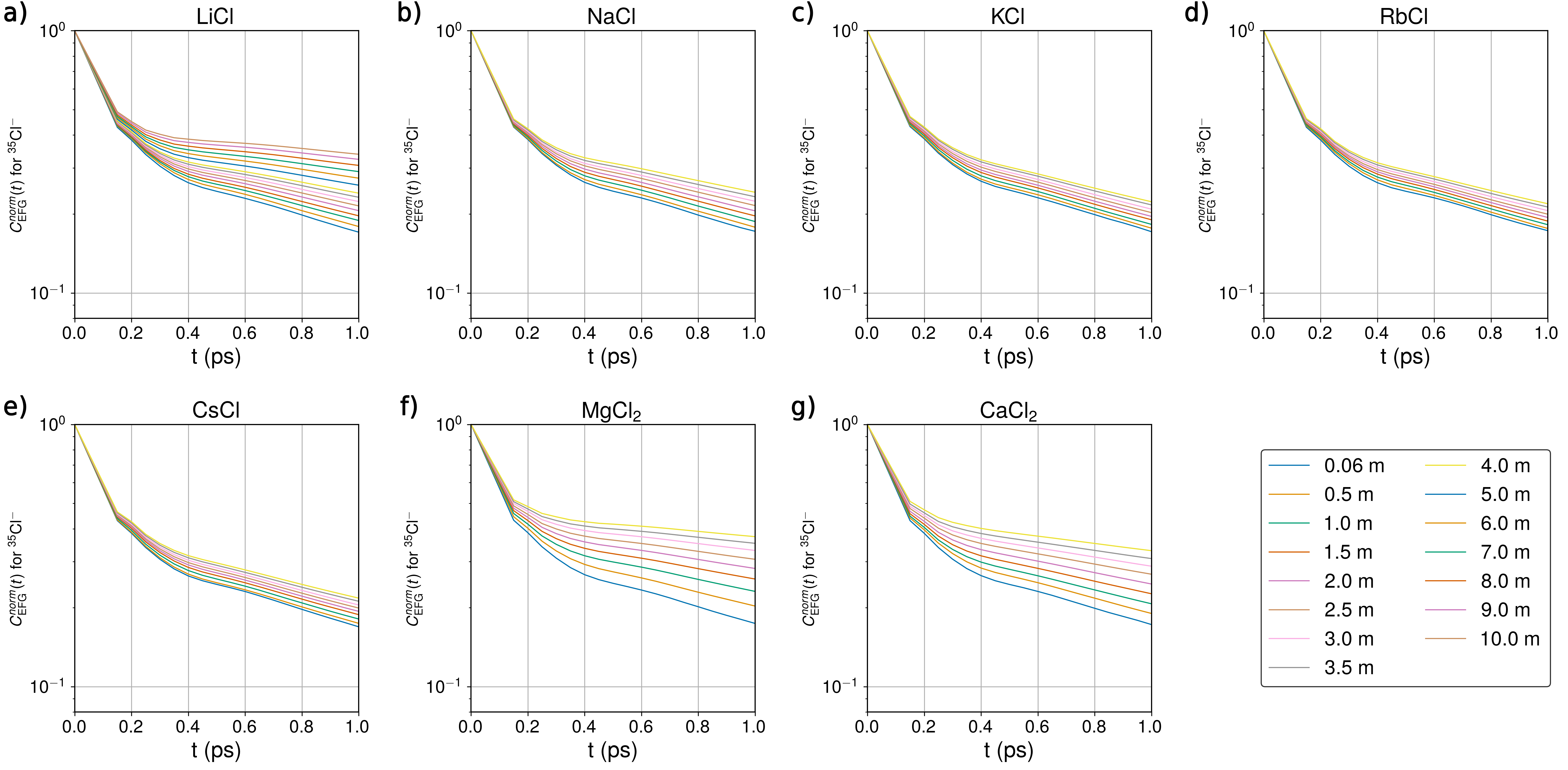}
    \caption{Normalized autocorrelatiocn function of the electric field gradient at the anion position in aqueous solutions of (a) LiCl, (b) NaCl, (c) KCl, (d) RbCl, (e) CsCl, (f) MgCl$_{2}$, and (g) CaCl$_{2}$, as a function of time, presented for various molalities indicated by the line colors.}
    \label{fig:ACf_Chlorine_zoom}
\end{figure}

\begin{figure}[ht!]
    \centering
    \includegraphics[width=\textwidth]{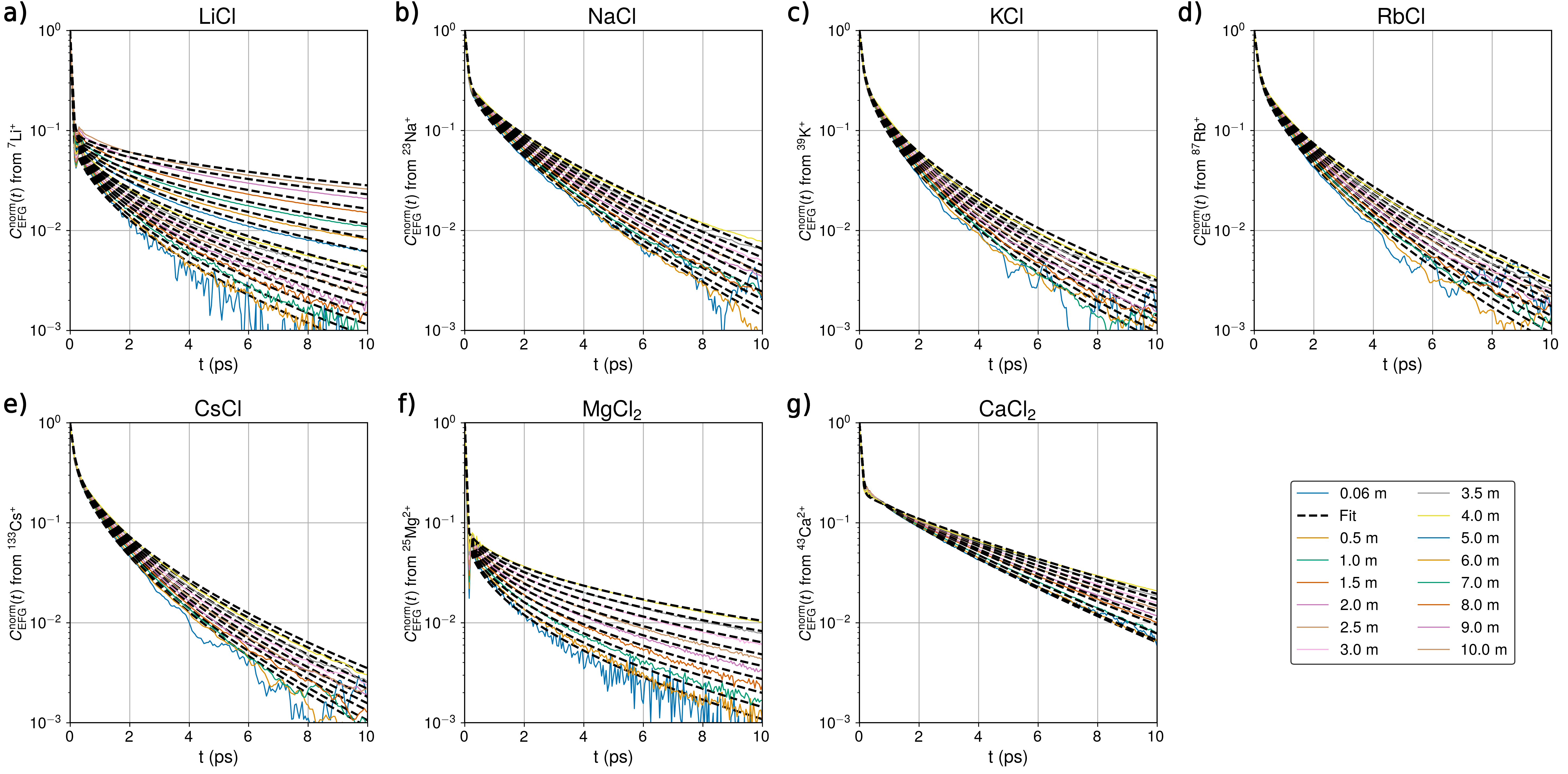}
    \caption{Normalized autocorrelation function of the electric field gradient at the cation position in aqueous solutions of (a) LiCl, (b) NaCl, (c) KCl, (d) RbCl, (e) CsCl, (f) MgCl$_{2}$, and (g) CaCl$_{2}$, as a function of time for various molalities. The black dashed line represents a fit obtained using the following function: $f(t) = (1 - \alpha _{s})e^{-t/\tau_{f}} + \alpha_{s} \cdot e^{-[t/\tau_{s}]^{\beta}}$.}
    \label{fig:ACF_IONS_fit}
\end{figure}

\begin{figure}[ht!]
    \centering
    \includegraphics[width=\textwidth]{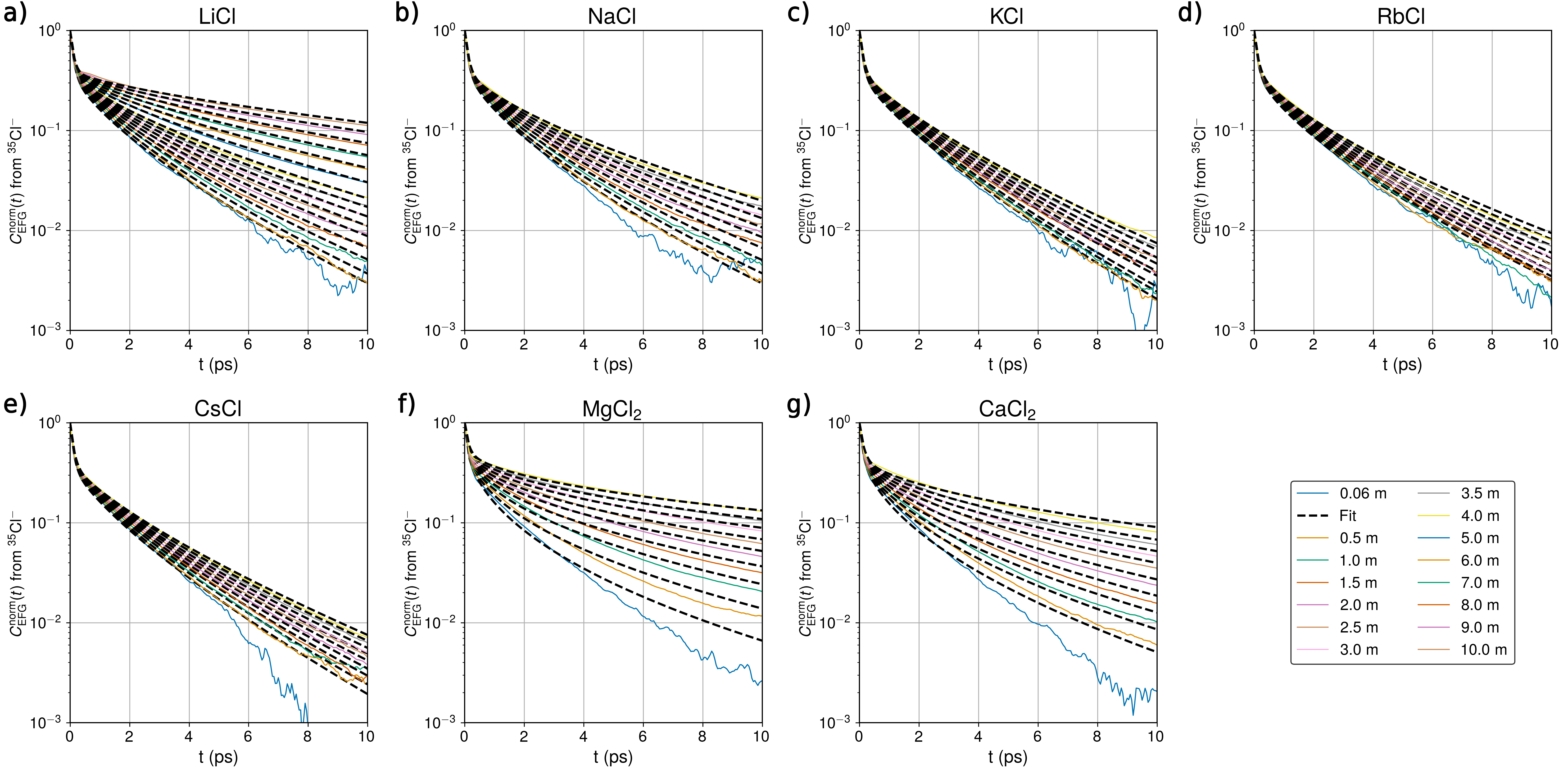}
    \caption{Normalized autocorrelation function of the electric field gradient at the anion position in aqueous solutions of (a) LiCl, (b) NaCl, (c) KCl, (d) RbCl, (e) CsCl, (f) MgCl$_{2}$, and (g) CaCl$_{2}$, as a function of time for various molalities. The black dashed line represents a fit obtained using the following function: $f(t) = (1 - \alpha _{s})e^{-t/\tau_{f}} + \alpha_{s} \cdot e^{-[t/\tau_{s}]^{\beta}}$.}
    \label{fig:ACF_IONS_fit_anion}
\end{figure}

\begin{figure}[ht!]
    \centering
    \includegraphics[width=\textwidth]{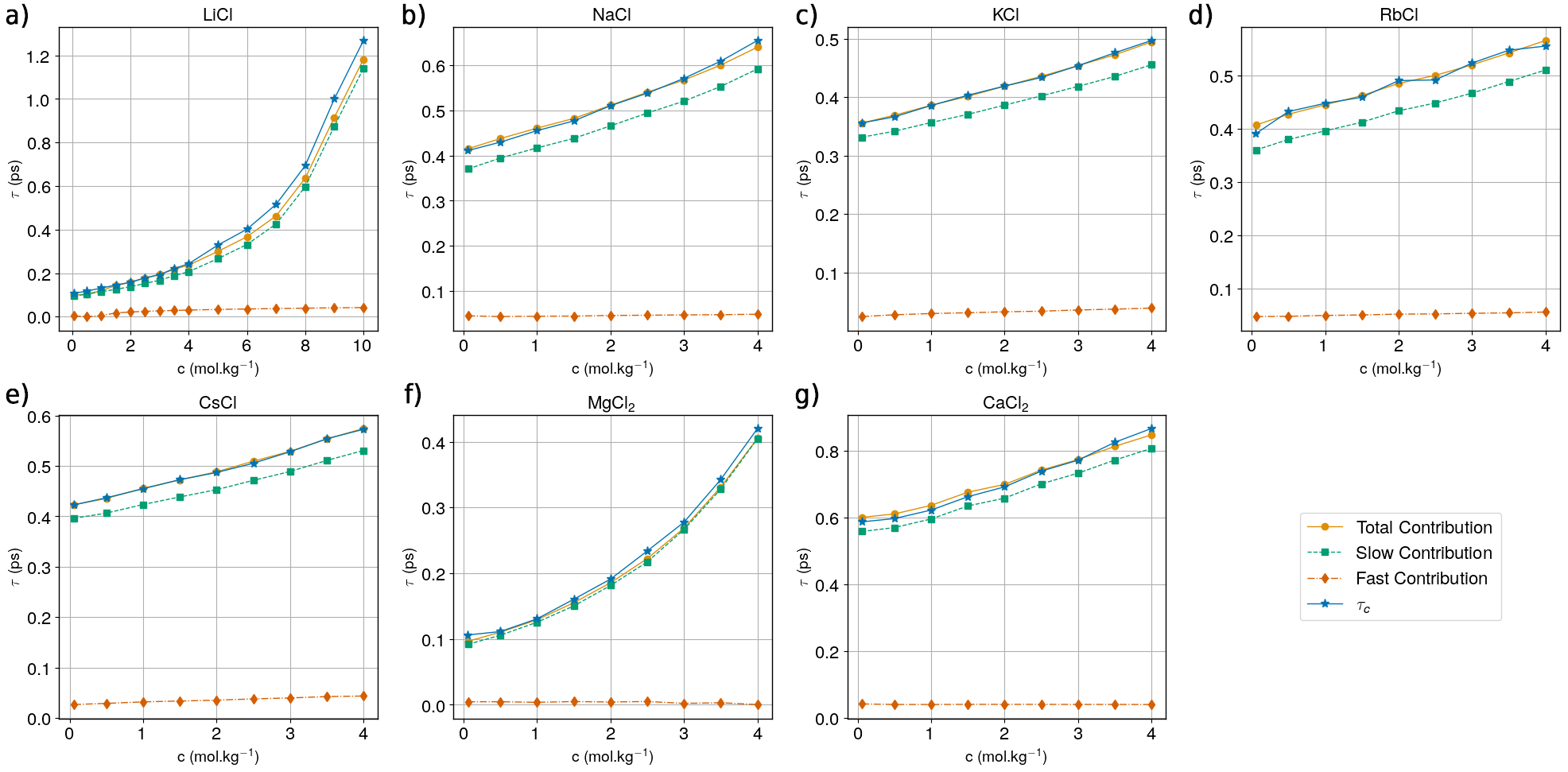}
    \caption{The contribution of the fast and slow modes to the total EFG decorrelation time $\tau_c$ at the cation position was evaluated by integrating the following function:  $f(t) = (1 - \alpha _{s})e^{-t/\tau_{f}} + \alpha_{s} \cdot e^{-[t/\tau_{s}]^{\beta}}$, and extracting the respective contributions of the fast and slow modes.}
    \label{fig:contrib_cation}
\end{figure}

\begin{figure}[ht!]
    \centering
    \includegraphics[width=\textwidth]{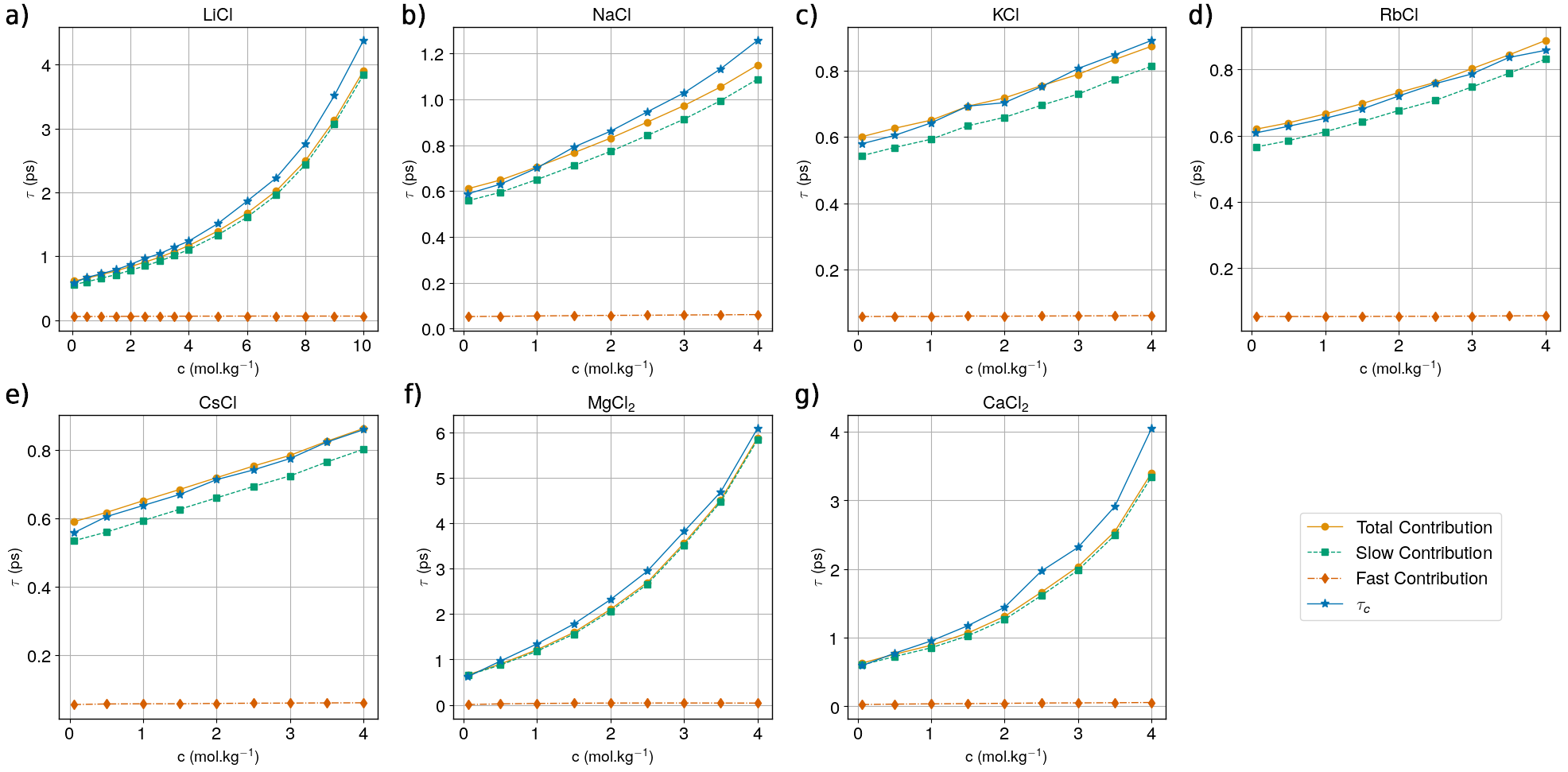}
    \caption{The contribution of the fast and slow modes to the total EFG decorrelation time $\tau_c$ at the anion position was evaluated by integrating the following function:  $f(t) = (1 - \alpha _{s})e^{-t/\tau_{f}} + \alpha_{s} \cdot e^{-[t/\tau_{s}]^{\beta}}$, and extracting the respective contributions of the fast and slow modes.}
    \label{fig:contrib_anion}
\end{figure}
\clearpage


\begin{table}[ht!]
    \centering
    \begin{tabular}{|c|c|c|c|c|c|c|c|}
        \toprule
        \hline
        \textbf{Molality (m)} & \textbf{LiCl} & \textbf{NaCl} & \textbf{KCl} & \textbf{RbCl} & \textbf{CsCl} & \textbf{MgCl$_2$} & \textbf{CaCl$_2$} \\
        \midrule
        \hline
        0.056 & 0.0189 & 16.58 & 14.26 & 611.29& 0.1630 & 8.86 & 1.2473 \\
        0.5   & 0.0208 & 17.31 & 14.60 & 676.29 & 0.1681 & 9.33 & 1.2663 \\
        1.0   & 0.0232 & 18.28 & 15.33 & 699.53 & 0.1749 & 10.98 & 1.3153 \\
        1.5   & 0.0258 & 19.08 & 15.99 & 715.46 & 0.1815 & 13.50 & 1.3949 \\
        2.0   & 0.0280 & 20.38 & 16.57 & 763.96 & 0.1868 & 16.11 & 1.4480 \\
        2.5   & 0.0316 & 21.42 & 17.13 & 765.10 & 0.1936 & 19.76 & 1.5413 \\
        3.0   & 0.0342 & 22.63 & 17.87 & 812.83 & 0.2023 & 23.50 & 1.5987 \\
        3.5   & 0.0393 & 24.08 & 18.70 & 849.40 & 0.2119 & 29.07 & 1.6994 \\
        4.0   & 0.0433 & 25.85 & 19.49 & 860.74 & 0.2189 & 35.74 & 1.7710 \\
        5.0   & 0.0587 & -       & -       & -        & -      & -       & -      \\
        6.0   & 0.0719 & -       & -       & -        & -      & -       & -      \\
        7.0   & 0.0927 & -       & -       & -        & -      & -       & -      \\
        8.0   & 0.1254 & -       & -       & -        & -      & -       & -      \\
        9.0   & 0.1822 & -       & -       & -        & -      & -       & -      \\
        10.0  & 0.2326 & -       & -       & -        & -      & -       & -      \\
        \hline
        \bottomrule
    \end{tabular}
    \caption{Computed relaxation rates ($1/T_1$ in s$^{-1}$) for the different cations in all considered aqueous electrolytes as a function of the salt concentration at 25$^\circ$C.}
    \label{tab:relaxation_rates_S1}
\end{table}

\begin{figure}[ht!]
    \centering
    \includegraphics[width=\textwidth]{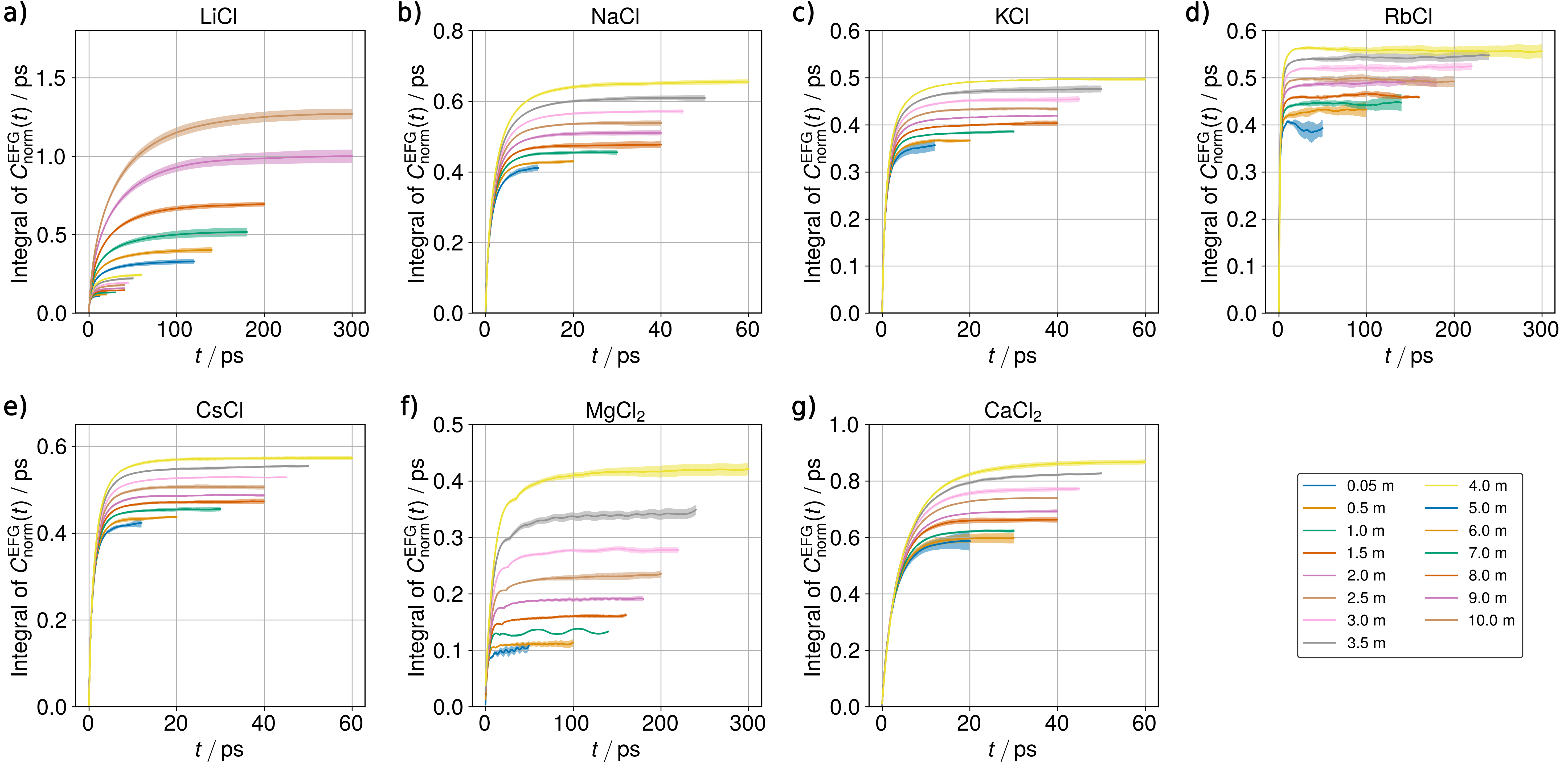}
    \caption{Time-dependent integrals of the normalized EFG ACFs at the position of the different cations in aqueous solutions of (a) LiCl, (b) NaCl, (c) KCl, (d) RbCl, (e) CsCl, (f) MgCl$_{2}$, and (g) CaCl$_{2}$, as a function of time, presented for various molalities indicated by the line colours. The shaded region indicates the standard error over different simulation runs. 
    }
    \label{fig:running_int_efg}
\end{figure}
\clearpage
III. D.

\begin{figure}[ht!]
    \centering
    \includegraphics[width=\textwidth]{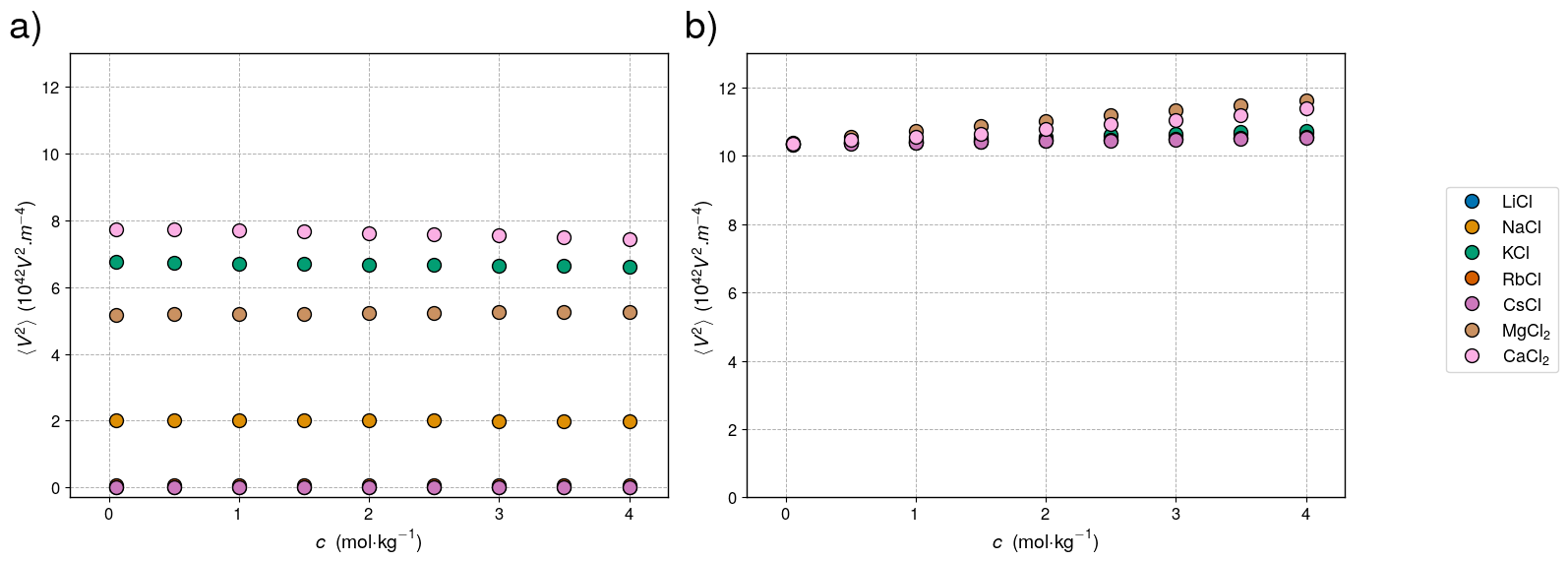}
    \caption{Variance of the EFG at the cation position a) and anion position b) $\langle V^{2} \rangle$ (including the effective Sternheimer factor) in aqueous (a) LiCl, (b) NaCl, (c) KCl, (d) RbCl, (e) CsCl, (f) MgCl$_{2}$, and (g) CaCl$_{2}$ solutions, as a function of molality.}
    \label{fig:EFG_variance_concentration}
\end{figure}

\clearpage
III. F. 1.

\begin{figure}[ht!]
    \centering
    \includegraphics[width=\textwidth]{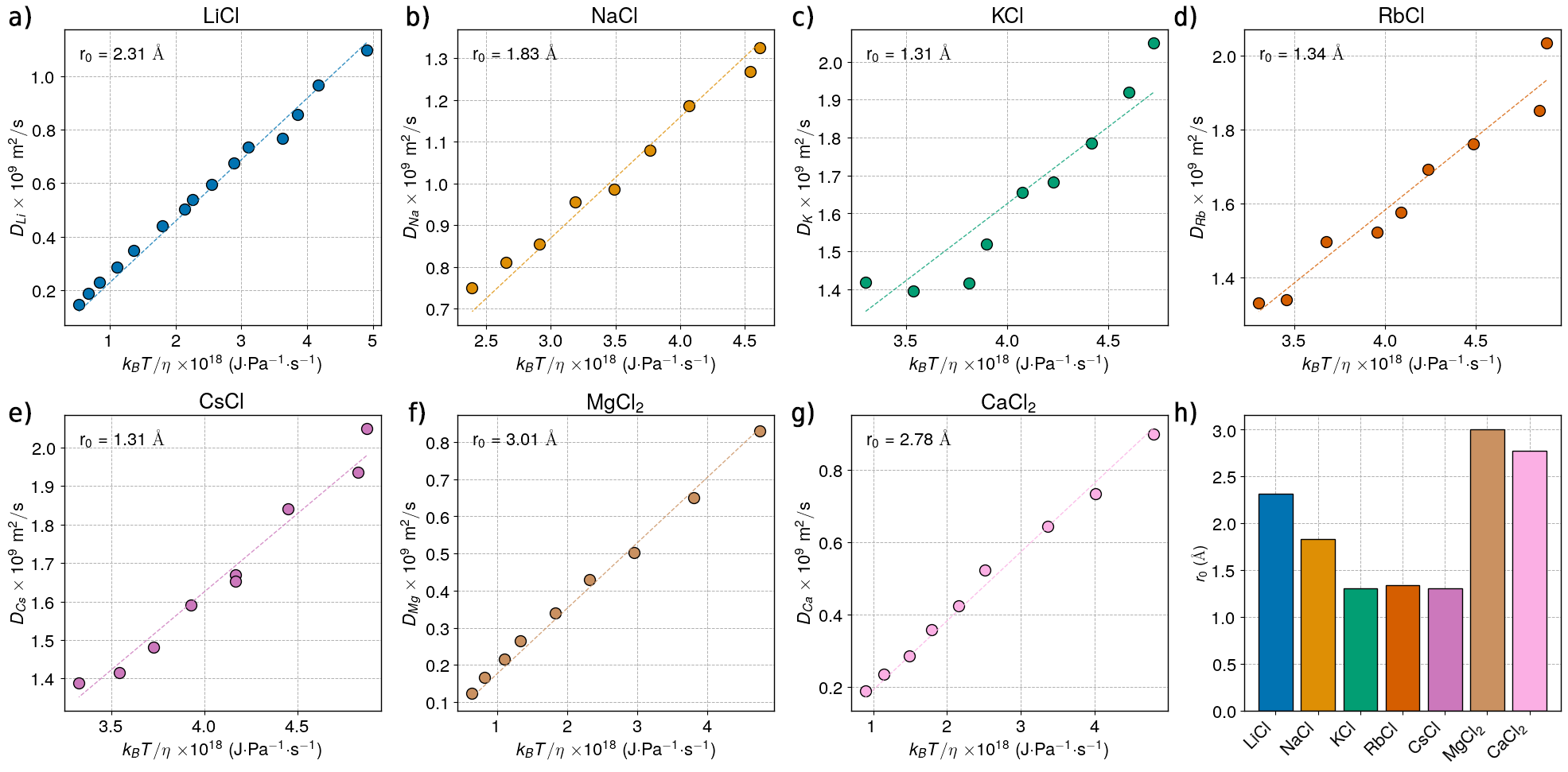}
    \caption{Cation diffusion coefficients for various systems as a function of $k_{B} T / \eta$. For each system, the dotted line represents the fit of the data to the relation $ D = k_{B}T/6 \pi \eta  r_{0}$, with $r_{0}$ indicated in the top left corner of each plot and gathered in the final bar plot. }
    \label{fig:hydroradii}
\end{figure}

\begin{figure}[ht!]
    \centering
    \includegraphics[width=\textwidth]{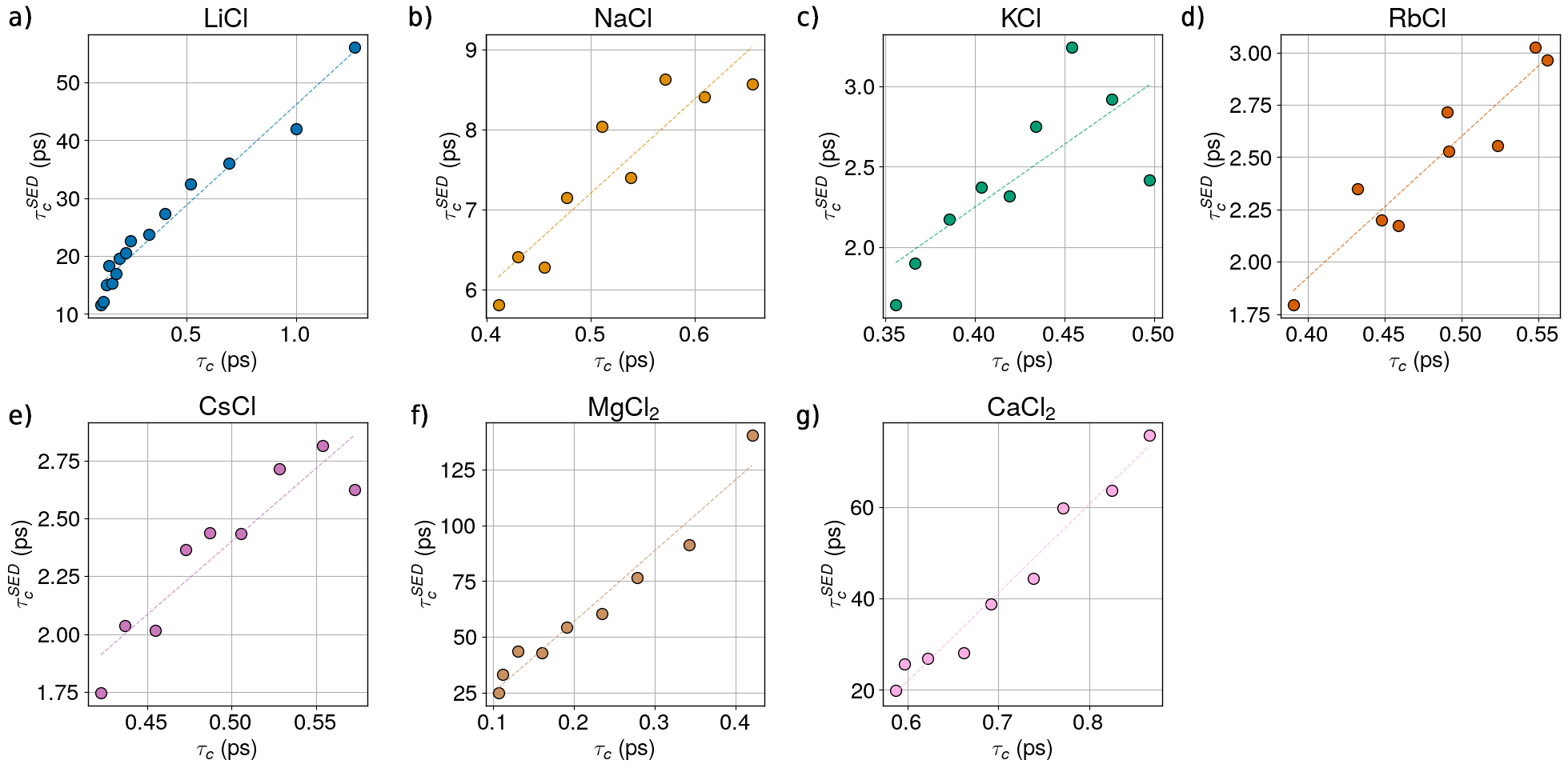}
    \caption{Stokes-Einstein-Debye correlation times $\tau_{SED}$ for different ions in each system, as calculated from Equation 11, are plotted against the electric field gradient (EFG) at the position of the cation from each system represented correlation times $\tau_{c}$ extracted from simulation data at various molalities. The grey dotted line represents a linear fit for visual guidance.}
    \label{fig:tautauSED}
\end{figure}

\begin{figure}[ht!]
    \centering
    \includegraphics[width=\textwidth]{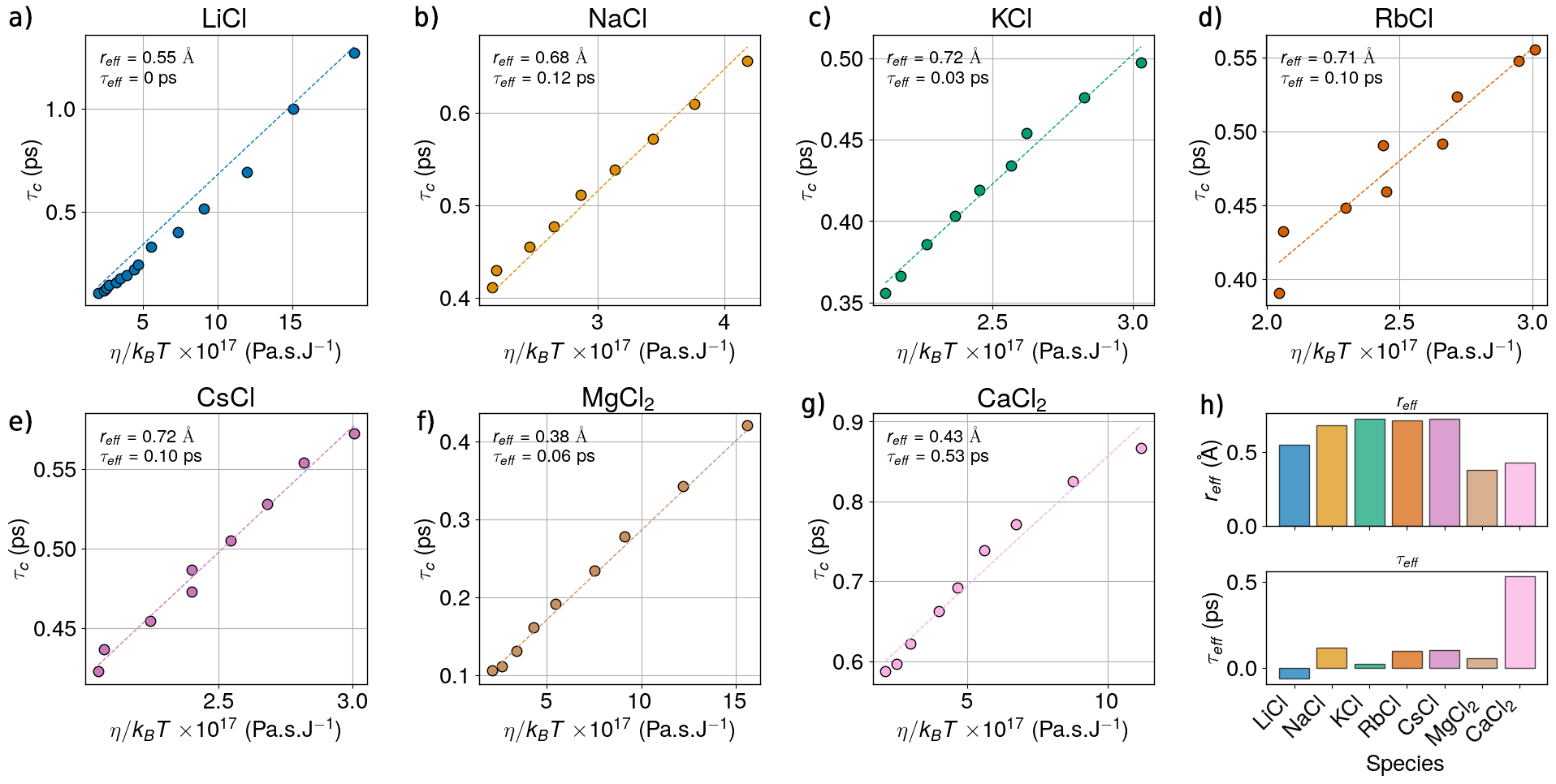}
    \caption{For the different ions of each system, $\tau_{c}$ is plotted versus $\eta/k_{B}T$ at different molalities. The dotted lines show the fit from the equation 14 with the best-fit parameters $r_{eff}$ and $\tau_{eff}$ indicated in the top corner left of each figure.}
    \label{fig:reff_tau}
\end{figure}

\clearpage
III. F. 2.

\begin{figure}[ht!]
    \centering
    \includegraphics[width=\textwidth]{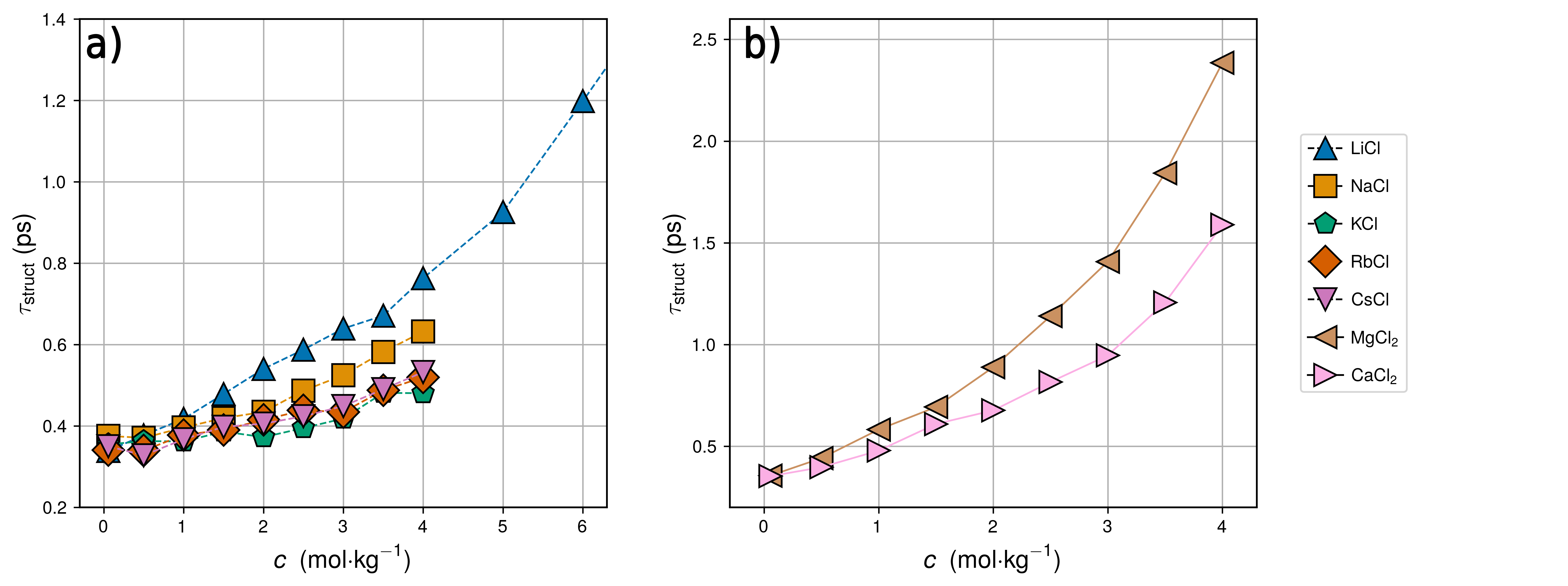}
    \caption{Structural relaxation time $\tau_{\rm struct}$ as a function of molality, for aqueous alkali metal chlorides (left) and alkaline earth chloride (right) solutions. The nature of the cation is indicated by the color and dashed lines are only guides for the eyes.
    }
    \label{fig:tau_struct}
\end{figure}

\end{document}